\documentclass[lettersize,journal]{IEEEtran}
\usepackage{amsmath,amsfonts}
\usepackage{algorithmic}
\usepackage{array}
\usepackage[caption=false,font=normalsize,labelfont=sf,textfont=sf]{subfig}
\usepackage{textcomp}
\usepackage{stfloats}
\usepackage{url}
\usepackage{verbatim}
\usepackage{graphicx}
\def\BibTeX{{\rm B\kern-.05em{\sc i\kern-.025em b}\kern-.08em
    T\kern-.1667em\lower.7ex\hbox{E}\kern-.125emX}}
\usepackage{balance}

\usepackage[style = ieee]{biblatex} 
\graphicspath{{Images/}}
\usepackage{algorithm}

\usepackage{import}
\usepackage{amssymb}
\usepackage{threeparttable}
\usepackage{booktabs}
\usepackage{tabularx}
\usepackage{mathtools}

\begin{document}

\title{Computationally-Efficient Synchrophasor Estimation: Delayed in-quadrature Interpolated DFT}

\author{César García-Veloso\thanks{Manuscript created August, 2022; The project that gave rise to these results received the support of a fellowship from ”la Caixa” Foundation (ID 100010434). The fellowship code is LCF/BQ/DR20/11790026.}\thanks{César García-Veloso and José María Maza-Ortega are with the Department of Electrical Engineering, University of Seville, 41092 Seville, Spain (e-mail: cgveloso@us.es; jmmaza@us.es).},~\IEEEmembership{Graduate Student Member, IEEE}, Mario Paolone\thanks{Mario Paolone is with the Distributed Electrical Systems Laboratory, École Polytechnique Fédérale de Lausanne, 1015 Lausanne, Switzerland (e-mail: mario.paolone@epfl.ch).},~\IEEEmembership{Fellow, IEEE}, José María Maza-Ortega,~\IEEEmembership{Member, IEEE}}


\newpage
"© 20xx IEEE. Personal use of this material is permitted. Permission from IEEE must be obtained for all other uses, in any current or future media, including reprinting/republishing this material for advertising or promotional purposes, creating new collective works, for resale or redistribution to servers or lists, or reuse of any copyrighted component of this work in other works."

Article submitted for consideration to IEEE Transactions on Instrumentation and Measurement. (Non peer reviewed)
\maketitle

\begin{abstract}
The paper proposes a synchropahsor estimation (SE) algorithm that leverages the use of a delayed in-quadrature complex signal to mitigate the self-interference of the fundamental tone. The estimator, which uses a three-point IpDFT combined with a three-cycle Hanning window, incorporates a new detection mechanism to iteratively estimate and remove the effects caused by interfering tones within the out-of-band interference (OOBI) range. The main feature of the method is its ability to detect interfering tones with an amplitude lower than that adopted by the IEC/IEEE Std. 60255-118, this detection being notably challenging. Furthermore, it simultaneously satisfies all the accuracy requirements for the P and M phasor measurement unit (PMU) performance classes, while offering a reduction in the total computational cost compared to other state-of-the-art techniques. Despite an increase in total memory requirements, a preliminary analysis reveals its suitability for implementation on embedded devices.
\end{abstract}


\section{Introduction}\label{doc_sec_1}
\IEEEPARstart{S}{ince} their introduction in the 1980s \cite{DeLaRee2010_Sync_PM_Applications_in_PS}, Phasor Measurement Units (PMUs) have evolved and are currently present in the power grids of all developed countries \cite{Martin2021_Synchrophasors_Point_the_Way}, used successfully in many applications, such as wide-area monitoring protection and control, model validation and state estimation \cite{702-TechnicalReport_PMU_CIGRE2017}. To ensure the inter-operability and compatibility of PMUs from different manufacturers, an industry standard, whose latest installment corresponds to the IEC/IEEE Std. 60255-118 \cite{PMU_Measurement_60255-118-1-2018}, has been in continuous development since 1992 \cite{Phadke2017_SynPhasorMeas_and_App}. The standard \cite{PMU_Measurement_60255-118-1-2018} defines a set of tests representative of the operating conditions of the power system, and the corresponding performance requirements that all PMUs must satisfy based on their class and reporting rate. Two different performance classes are defined in \cite{PMU_Measurement_60255-118-1-2018}: class P, for applications requiring a fast response, and class M, for those requiring high accuracy and rejection of interharmonics or subharmonics.

Over the years many synchrophasor estimation (SE) techniques have been explored in the literature such as wavelets \cite{Ren2011FreqPhasorsRWavelet}, Prony \cite{Serna2013SynchrophasorEstimationUsingPronysMethod}, Taylor-Fourier \cite{Serna2007DynamicPhasors,Platas-Garza2010TaylorFourierFilter,Bertocco2015_CSTF_SyncEst}, Shanks \cite{Munoz2008_ShanksMethodDynPhEst}, Kalman filtering \cite{Serna2011TaylorKKalmanFilters, Liu2012ModifiedTaylorKalman,Serna2012TaylorKalmanFourier} and adaptive filters \cite{Roscoe2013PandMClassPMUAlgoUsingAdaptiveCascadedFilters}, among others. 
Most commercial PMUs employ DFT-based SE algorithms \cite{Kamwa2014_WFRA_PMU_Algos}, which can produce accurate estimates by simply processing a few DFT bins \cite{Belega2014_Freq_est_sinu_3pIpDFT_ImRej}. However, two main limitations compromise the performance of DFT-based techniques: aliasing and spectral leakage \cite{Reljin2007ExtFlatTopWind}. Although the first can be easily solved by increasing the sampling rate or adopting an anti-aliasing filter, spectral leakage, which includes long- and short-range interferences, requires more elaborate solutions \cite{Reljin2007ExtFlatTopWind}. Long-range leakage refers to the mutual interference caused by all the tones that make up the signal spectrum, while short-range leakage, also known as scalloping or picket fence effect, is the error caused by the displacement of the maximum bin. Long-range leakage can be mitigated by windowing \cite{Harris1978WindowsforHarmonicAnalysisDFT}, while interpolation of DFT bins \cite{Jain1979IpFFT} addresses short-range leakage. Combining both windowing and interpolation, as proposed in \cite{Grandke1983InterpolationAlgorithmsforDFTs}, represents the foundation of modern interpolated DFT (IpDFT) techniques. To satisfy the underlying assumption that the signal parameters are fixed within the observation window, and to comply with the time and latency requirements set by \cite{PMU_Measurement_60255-118-1-2018}, intervals of just a few nominal cycles are generally selected \cite{Romano2014EnhancedInterpolatedDFT, Derviskadic2018i-IpDFTforSynchrophasorEstimation, Frigo2019ReducedLekageSynPhEstimation}. In turn, this causes a coarse frequency resolution which, in the case of a real-valued power system signal whose main tone nominal values fall close to DC, results in the proximity between the positive and negative images. Their mutual interaction, also known as self-interference, constitutes the main source of error in the estimation process \cite{Frigo2019ReducedLekageSynPhEstimation}.

Different approaches have been explored to deal with negative frequency infiltration \cite{Agrez2002_WM_IpDFT, Romano2014EnhancedInterpolatedDFT, Derviskadic2018i-IpDFTforSynchrophasorEstimation,Frigo2019ReducedLekageSynPhEstimation,Zhan2018AClarkeTransfBasedDFTPhasorandFreqAlgorithm,Belega2014_Sine_est_IpDFT_cos_wind,Belega2014_Freq_est_sinu_3pIpDFT_ImRej,Belega2021_accu_param_est_sines}. \cite{Agrez2002_WM_IpDFT} introduced a multi-point weighted IpDFT that reduces the effects of long-range spectral leakage. \cite{Belega2014_Freq_est_sinu_3pIpDFT_ImRej} employed a method based on a novel three-point weighted IpDFT combined with the use of maximum sidelobe decay (MSD) windows, which showed a great rejection of interference from the negative image. \cite{Belega2014_Sine_est_IpDFT_cos_wind} developed new cosine windows, called Maximum Image interference Rejection with Rapid Sidelobe Decay rate (MIR-RSD) windows, which showed a high rejection of self-interference and long-range leakage from other narrowband disturbances. \cite{Romano2014EnhancedInterpolatedDFT} presented an algorithm (e-IpDFT) that iteratively approximated and compensated for the effects of self-interference by taking advantage of the symmetry of the DFT spectrum with respect to the DC component. The method was further extended in \cite{Derviskadic2018i-IpDFTforSynchrophasorEstimation} to further compensate for an additional generic interfering tone. The resulting method, the so-called i-IpDFT, was proven to meet the requirements for the P and M classes. Both e-IpDFT and i-IpDFT were proved to be computationally efficient and experimentally deployed and validated on industry-grade field-programmable gate arrays (FPGAs) \cite{Romano2014_mSDFT_eIpDFT, Derviskadic2020_iIpDFT_FPGA}. The same idea of estimating and compensating for the effects of self-interference and long-range leakage from additional tones presented in \cite{Derviskadic2018i-IpDFTforSynchrophasorEstimation} is also used in 
\cite{Belega2021_accu_param_est_sines}. While \cite{Derviskadic2018i-IpDFTforSynchrophasorEstimation} uses an iterative formulation, \cite{Belega2021_accu_param_est_sines} derives analytical expressions to achieve faster convergence. However, \cite{Belega2021_accu_param_est_sines} does not account for subharmonic or interharmonic components.

A different approach to mitigate negative image infiltration is to generate a complex signal from the real one.  HT-IpDFT \cite{Frigo2019ReducedLekageSynPhEstimation} employs a Hilbert filter to approximate the analytical signal and suppress the negative spectrum. Although the method reduces the computational complexity of the iterative process compared to \cite{Derviskadic2018i-IpDFTforSynchrophasorEstimation}, it does not meet the combined requirements of classes P and M for harmonic distortion (1\%) and phase step tests in \cite{PMU_Measurement_60255-118-1-2018}.  Another suitable complex signal to cancel the negative image is the one defined by the signal itself and its imaginary in-quadrature signal. Many quadrature signal generation (QSG) techniques exist and have been used in other applications, such as single-phase phase-locked loops (PLLs) \cite{Golestan2017_1PhPLL_Review}. In \cite{Zhan2018AClarkeTransfBasedDFTPhasorandFreqAlgorithm} an in-quadrature signal is obtained by applying the Clarke transform to a three-phase signal generated by buffering and shifting a noise-filtered version of the measured single-phase signal. The complex signal is further enhanced with a weighted least-squares Taylor-Fourier (WLS-TF) filter to match the magnitudes of both in-quadrature components and is used with a Blackman window filter.

This paper proposes a SE algorithm that takes advantage of the use of a delayed in-quadrature complex signal generated in the time domain to mitigate the self-interference of the fundamental tone. The estimator, named Time-Delay IpDFT (TD-IpDFT), uses a three-point IpDFT combined with a Hanning window and incorporates a new detection mechanism to iteratively estimate and remove the effects caused by interfering tones equal to or greater than 5\% within the out-of-band interference (OOBI) range. It simultaneously satisfies all accuracy requirements for the P and M classes, while offering a reduction in the total computational cost compared to the i-IpDFT \cite{Derviskadic2018i-IpDFTforSynchrophasorEstimation,Derviskadic2020_iIpDFT_FPGA}.

The paper is structured as follows. Section \ref{doc_sec_2} reviews the fundamentals of IpDFT-based SE techniques. Section \ref{doc_sec_3} describes the theoretical basis of how a complex signal consisting of in-quadrature components allows the suppression of the negative image spectrum. Section \ref{doc_sec_4} presents the structure, formulation, and adjustment of the parameters of the TD-IpDFT SE algorithm. Section \ref{doc_sec_5} presents the complete characterization of the algorithm for P- and M-class PMUs indicated by \cite{PMU_Measurement_60255-118-1-2018}. Section \ref{doc_sec_6} discusses and compares the results with those obtained by other state-of-the-art SE methods. Finally, \ref{doc_sec_7} summarizes the main findings and gives closing remarks.

\section{SE IpDFT Techniques: Fundamentals}\label{doc_sec_2}
We refer to a formulation that considers a Hanning window function, which offers a good compromise between sidelobe decay and mainlobe width \cite{Grandke1983InterpolationAlgorithmsforDFTs}, and a three-point DFT interpolation scheme, as it reduces the effects of long-range leakage \cite{Agrez2002_WM_IpDFT}. This configuration, already adopted in \cite{Derviskadic2018i-IpDFTforSynchrophasorEstimation, Frigo2019ReducedLekageSynPhEstimation}, is also used for the proposed TD-IpDFT.
\subsection{Three-point (3p) IpDFT based on the Hanning window}
Consider a discrete single-tone steady-state signal $x(n)$:
\begin{equation}
\label{eq_sing_tone_SS}
x(n) = A_{0} \cos{(2\pi f_{0} n T_{s} + \varphi_{0})}, \quad n\in \mathbb{N}
\end{equation}
where $T_{s}$ represents the sampling period and $A_{0}$, $f_{0}$ and $\varphi_{0}$ the fundamental amplitude, frequency and initial phase, respectively.
If a window of $N$ consecutive samples is selected, the DFT spectrum of the signal $X(k)$ is given by (\ref{eq_DFT_def}), where $w(n)$ is the discrete windowing function and $B = \sum_{n=0}^{N-1}w(n)$ the DFT normalization factor.
\begin{equation}
\label{eq_DFT_def}
X(k) = \frac{1}{B}\sum_{n=0}^{N-1}w(n)x(n)e^{-j2\pi k n/N}, \quad k\in[0,N-1]
\end{equation}
According to the convolution theorem, a conjugate spectrum is obtained, where each spectral component $X(k)$ is the result of the combined contributions of a positive and negative image of the fundamental tone, each being the scaled and rotated versions of the DFT of the window function $W(k)$ shifted at $\pm f{_0}/\Delta_f$\footnote{In the case of a real multi-tone signal its spectrum will be result of the combined contributions from the positive and negative images of each individual tone.}:
\begin{equation}
\label{eq_DFT_pos_neg}
X(k) = \underbracket[.5pt]{V_{0}W(k-f_{0}/\Delta_{f})}_{\text{Positive Image}} + \underbracket[.5pt]{V_{0}^{*}W(k+f_{0}/\Delta_{f})}_{\text{Negative Image}}
\end{equation}
where $V_{0} = A_{0} e^{j\varphi_{0}}$. The distance between positive and negative images depends on the frequency resolution $\Delta_{f}$, which is the reciprocal of the length of the selected window $T = NT_{s}$. For the Hanning window, $W(k)$ corresponds to:
\begin{equation}
\label{eq_Hann_wind}
W_{H}(k) =  0.5W_{R}(k) -0.25(W_{R}(k-1) + W_{R}(k+1))
\end{equation}
where $W_{R}(k)$ is the DFT of the rectangular window, also known as the Dirichlet kernel:
\begin{equation}
\label{eq_rect_wind}
W_{R}(k) = e^{-j\pi k (N-1)/N} \frac{\sin{(\pi k)}}{\sin{(\pi k/N)}}, \quad k\in[0,N-1]
\end{equation}
Given a set of estimates of the signal parameters $(\hat{f},\hat{A},\hat{\varphi})$ and knowing the windowing function adopted, the spectral contributions of positive and negative images can be estimated by:
\begin{equation}
\label{eq_wf_def}
\hat{X}_{H\pm}(k) =  \hat{A} e^{\pm j\hat{\varphi}} W_{H}(k\mp\hat{f}/\Delta_{f})
\end{equation}
For the general case of incoherent sampling\footnote{Incoherent sampling refers to the adoption of a window length ($T$) which does not contain an integer number of fundamental periods ($1/f_{0}$), i.e., $\delta\neq0$.}, the fundamental tone of the signal falls between subsequent bins. Thus, the signal frequency can be expressed as
\begin{equation}
\label{eq_delta_def}
f_{0} = (k_{m} + \delta) \Delta_{f}, \quad \delta \in [-0.5,0.5)
\end{equation}
where $k_{m}$ is the index of the highest bin and $\delta$ a fractional correction term that, for the Hanning window ($\delta_{H}$), can be analytically calculated by interpolating the three highest DFT bins as \cite{Agrez2002_WM_IpDFT}:
\begin{equation}
\label{eq_delta_hann}
\delta_{H} = 2\varepsilon \frac{|X_{H}(k_{m}+\varepsilon)|-|X_{H}(k_{m}-\varepsilon)|}{|X_{H}(k_{m}-\varepsilon)|+2|X_{H}(k_{m})| +|X_{H}(k_{m}+\varepsilon)|}
\end{equation}
where $\varepsilon=\pm 1$ if $|X_{H}(k_{m}+1)| \gtrless |X_{H}(k_{m}-1)|$.

With $\delta_{H}$ determined, the amplitude and phase of the fundamental tone are given by: 
\begin{subequations}\label{eq_amp_ph_hann}
\begin{align}
A_{0_{H}} &= 2|X_{H}(k_{m})| \left|\frac{\pi\delta_{H}}{\sin{(\pi\delta_{H})}}\right||\delta_{H}^{2}-1| \label{eq_amp_hann}\\
\varphi_{0_{H}} &= \angle{X_{H}(k_{m})}-\pi\delta_{H}\label{eq_ph_hann}
\end{align}
\end{subequations}

\section{Negative Fundamental Image Suppression}\label{doc_sec_3}
\subsection{Delayed in-quadrature technique}
\begin{figure}[!t]
\centering
\includegraphics[width=\linewidth]{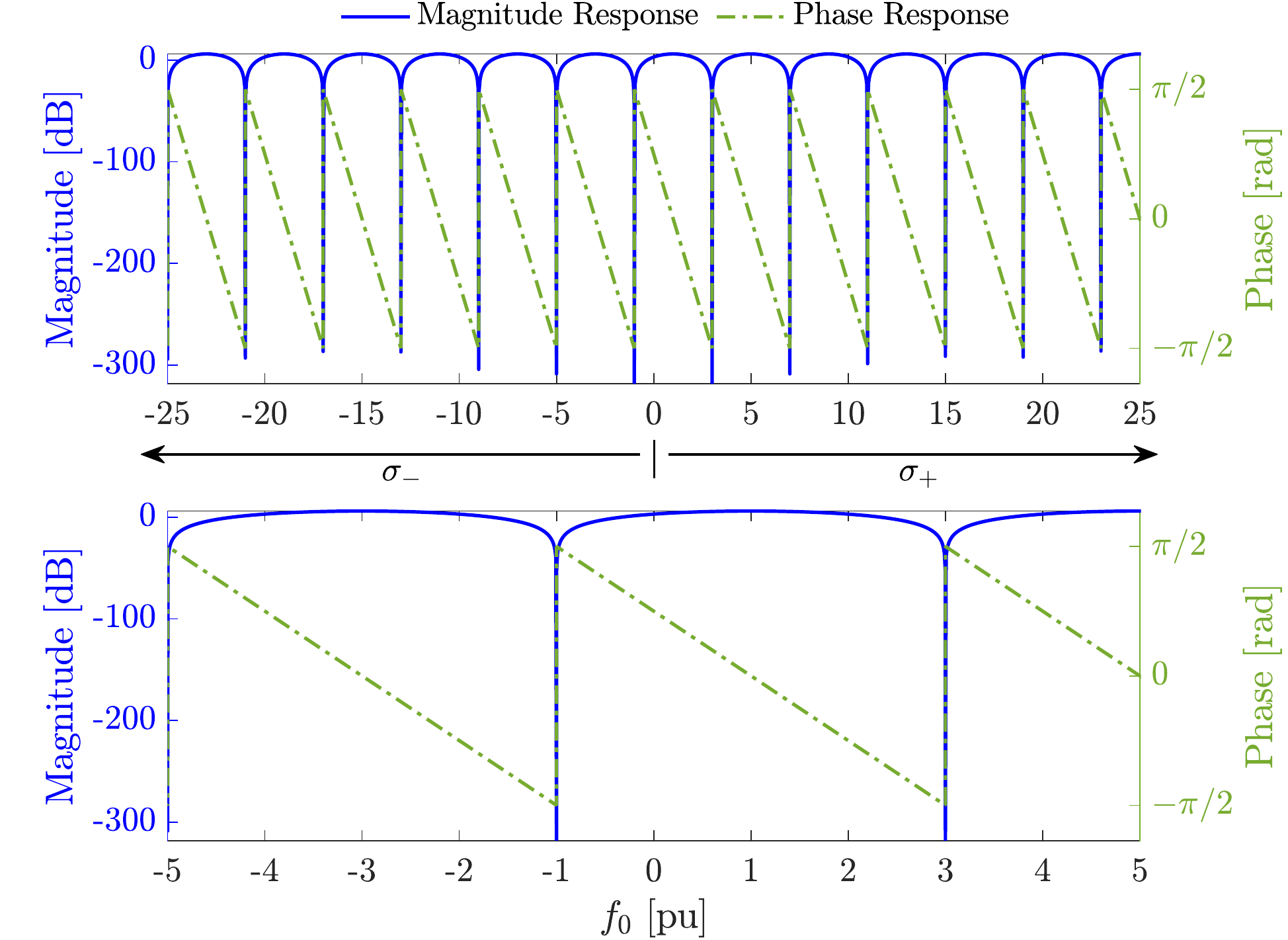}
\caption{Frequency response of a 'filtered' delayed in-quadrature complex signal $\bar{y}(n)$ generated considering a frequency $f$ from a signal $y(n)$ with normalized frequency $f_{0}[\text{pu}] = f_{0}/f$. Wide view (top) and zoomed (bottom).}
\label{fig_TD_Filter_resp}
\end{figure}

Consider a complex discrete single-tone steady-state signal $\bar{y}(n)$, with components $y(n)$ and $y(n-d_{\theta})$, $\theta$ being the resulting phase shift that delay $d_{\theta}$ $(d_{\theta} \in \mathbb{N})$ introduces at the corresponding frequency $f_{0}$:

\begin{equation}
\label{eq_yphshift_cont_trig}
\bar{y}(n) = \underbracket[.5pt]{A_{0}\cos{(\omega_{0} n T_{s})}}_{y(n)} + j\underbracket[.5pt]{A_{0}\cos{(\omega_{0} n T_{s} - \theta})}_{y(n-d_{\theta})}, \; n\in \mathbb{N}
\end{equation}where $\omega_{0} = 2\pi f_{0}$ is the angular frequency\footnote{Although for simplicity the initial angle has been set to 0, the same result holds if considered.} and $\theta = \omega_{0}d_{\theta}T_{s}$. If (\ref{eq_yphshift_cont_trig}) is expressed in terms of complex exponentials and the positive and negative frequency components are grouped, the following expression is obtained:

\begin{equation}
\label{eq_yphshift_cont_cexp}
\bar{y}(n) = \underbracket[.5pt]{\frac{1}{2}A_{0} e^{j(\omega_{0} n T_{s})}\sigma_{\text{+}}}_{\text{Positive Frequency}} + \underbracket[.5pt]{\frac{1}{2}A_{0} e^{-j(\omega_{0} n T_{s})}\sigma_{\text{--}}}_{\text{Negative Frequency}}, \; n\in \mathbb{N}
\end{equation}where $\sigma_{\text{+}}$ and $\sigma_{\text{--}}$ are the complex positive and negative delay gains and correspond to:
\begin{subequations}\label{eq_sigmas_pn}
\begin{align}
\sigma_{\text{+}}=  1 + e^{j (\pi/2 -\theta)}\label{eq_sigmap}\\
\sigma_{\text{--}}=  1 + e^{j (\pi/2 +\theta)}\label{eq_sigman}
\end{align}
\end{subequations}
The complex signal $\bar{y}(n)$ can be seen as a 'filtered' version of $y(n)$, where the amplitude and phase of each positive and negative frequency component are altered respectively by (\ref{eq_sigmap}) and (\ref{eq_sigman}) based on the relative angular shift $\theta$. Let $\lfloor...\rceil$ be the round-to-the-nearest integer function, rounding the equidistant values away from zero to the integer with a larger magnitude. Fig \ref{fig_TD_Filter_resp} shows the frequency response of a 'filtered' delayed in-quadrature complex signal $\bar{y}(n)$ generated considering a frequency $f$ from a signal $y(n)$ with normalized frequency $f_{0}[\text{pu}] = f_{0}/f$. The magnitude and phase response for each positive and negative frequency component are given, respectively, by $|\sigma_{\text{+}}|,\angle{\sigma_{\text{+}}}$ and $|\sigma_{\text{--}}|,\angle{\sigma_{\text{--}}}$ with $d_{\theta} = \left \lfloor f_{s}/(4 f) \right \rceil$. The frequency response is 4 pu periodic and is centered on the frequency considered for in-quadrature generation $f$ (1 pu). If there is no discrepancy between the signal frequency ($f_{0}$) and that used for the generation of the in-quadrature component, i.e. ($f_{0} = f \rightarrow f_{0}[\text{pu}]=1$), a perfect in-quadrature signal is obtained, resulting in cancelation of the negative image ($\sigma_{\text{--}}$ = 0) and doubling of the positive one ($\sigma_{\text{+}}$ = 2). Otherwise, if the signal frequency falls in its vicinity ($f_{0} \simeq f \rightarrow f_{0}[\text{pu}] \simeq 1$), still a significant mitigation of the negative image occurs, but the effects can only be quantified by (\ref{eq_sigmas_pn}) once the signal frequency has been estimated. 

\subsection{Adopted delayed structure}
A delayed in-quadrature complex signal generation technique is adopted. Conceptually, an in-quadrature component, given a sampled single-tone steady-state signal ($x(n)$) as defined in (\ref{eq_sing_tone_SS}), can be generated by delaying it by $\left \lfloor f_{s}/(4f_{0})\right \rceil$. Since $f_{0}$ is not known in advance, a two-step delayed in-quadrature complex signal generation method based on the IpDFT, the so-called Time-Delay QSG (TD-QSG), is proposed. The method is described in Algorithm \ref{alg_TDQSG}, where the functions $\mathtt{DFT}$ and $\mathtt{IpDFT}$ refer respectively to (\ref{eq_DFT_def}) and (\ref{eq_delta_def})-(\ref{eq_amp_ph_hann}).

First, an initial approximation of the in-quadrature complex signal ($\bar{x}_{o}(n)$) is generated delaying the sampled signal ($x(n)$) by $d_{0}$ samples. These correspond to the rounded delay given by the nominal frequency $f_{n}$ (lines 1-2). The spectrum of the complex signal is then calculated and windowed in the frequency domain considering a three-cycle Hanning window\footnote{This has been found to be the shortest window length required to detect and remove all tones defined by the out-of-band interference test.} (lines 3-4). A three-point IpDFT is then applied to just obtain a refined estimate of the signal frequency ($\hat{f}_{0}$) (line 5), which is subsequently used to obtain a refined delay $d_{f}$ and in-quadrature complex signal ($\bar{x}_{f}(n)$) (lines 6-7).



\begin{algorithm}
\caption{TD-QSG Algorithm}
\label{alg_TDQSG}
\begin{algorithmic}[1]
\REQUIRE $[x(n)]$
\STATE $d_{0} = \left \lfloor \frac{f_{s}}{4 f_{n}} \right \rceil$
\STATE $\bar{x}_{o}(n) = x(n) + j x(n-d_{0})$
\STATE $X_{o}(k) = \mathtt{DFT}[\bar{x}_{o}(n)]$
\STATE $ X_{o_{H}}(k)  =  0.5X_{o}(k) -0.25(X_{o}(k-1) + X_{o}(k+1))$
\STATE $\{\hat{f_{0}}\} = \mathtt{IpDFT}[X_{o_{H}}(k)]$
\STATE $d_{f} = \left \lfloor \frac{f_{s}}{4\hat{f_{0}}} \right \rceil$
\STATE $\bar{x}_{f}(n) = x(n) + j x(n-d_{f})$
\ENSURE $\{\bar{x}_{f}(n)\}$
\end{algorithmic}
\end{algorithm}


\section{TD-IpDFT Technique}\label{doc_sec_4}
This section describes the TD-IpDFT algorithm by providing: (i) a description of its structure and formulation (Section \ref{doc_sec_4_sub1}); (ii) a novel trigger mechanism for OOBI identification and removal; (iii) a sensitivity analysis to choose its most suitable parameters (Section \ref{doc_sec_4_sub2}), and (iv) a preliminary analysis of its computational complexity (Section \ref{doc_sec_4_sub3}). The proposed method is an improvement over the i-IpDFT proposed in \cite{Derviskadic2018i-IpDFTforSynchrophasorEstimation,Derviskadic2020_iIpDFT_FPGA} and its aim is to match the performance of the i-IpDFT while presenting a lower computational cost by using a delayed in-quadrature complex signal to mitigate the effects of self-interference. 

\subsection{Structure and Formulation}\label{doc_sec_4_sub1}
\begin{figure}[!t]
\centering
\includegraphics[width=\linewidth]{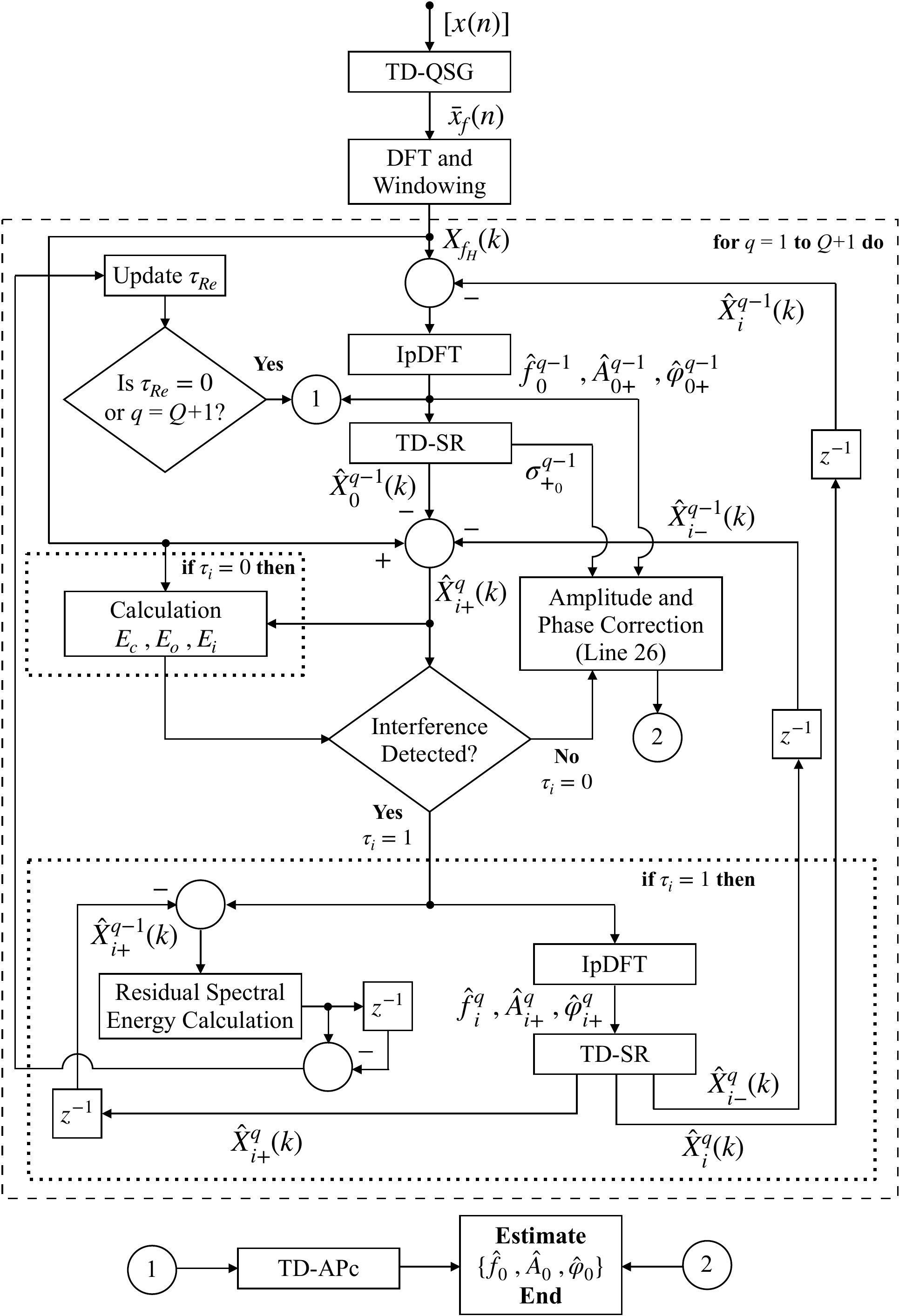}
\caption{Diagram of the TD-IpDFT Algorithm. The process is looped Q + 1 times instead of Q as in the pseudocode in Algorithm \ref{alg_TDIpDFT} since the $\mathtt{IpDFT}$ function on line 4 is nested within the loop.}
\label{fig_FlowChart_Algo}
\end{figure}

\begin{figure*}[!t]
\centering
\includegraphics[width=\linewidth]{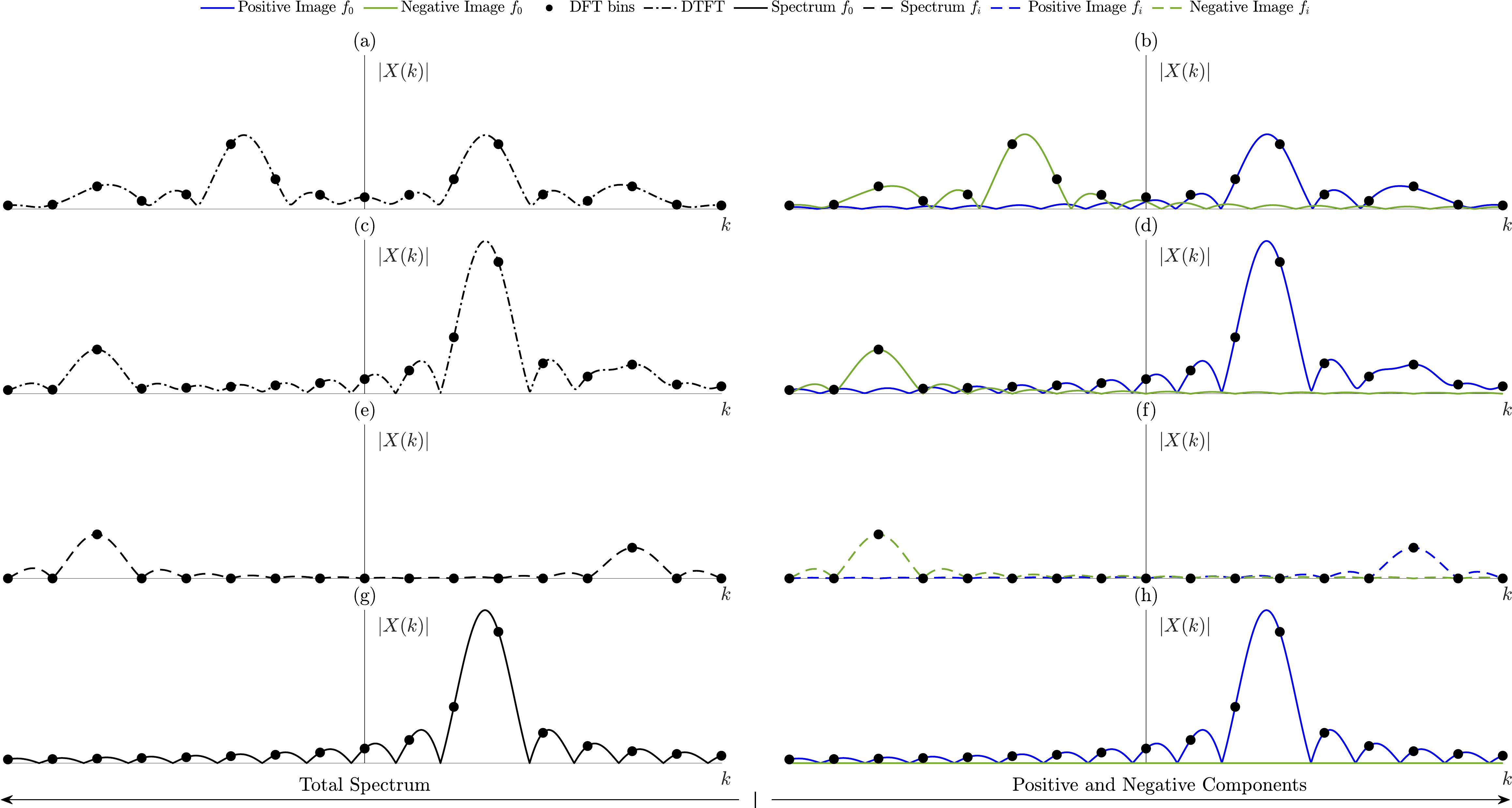}
\caption{Spectra at different steps of Algorithm \ref{alg_TDIpDFT} for a discrete signal $x(n)$ affected by additive white Gaussian noise with a signal-to-noise ratio equal to 80dB $[x(n) = A_{0} \cos{(2\pi f_{0} n T_{s})} + A_{i} \cos{(2\pi f_{i} n T_{s})}, \; n\in \mathbb{N}$; where $A_{0} = 1$, $f_{0} = 45$ Hz, $A_{i} = 0.35$ and $f_{i} = 100$ Hz$]$ : (a)-(b) $X_{0_{H}}(k)$ in line 1 (line 4 of Algorithm \ref{alg_TDQSG}); (c)-(d) $X_{f_{H}}(k)$ in line 3; (e) $X_{f_{H}}(k) - \hat{X}_{0}^{q-1}(k)$ within line 9; (f) $\hat{X}_{i}^{q}(k)$ in line 23; (g) $X_{f_{H}}(k) - \hat{X}_{i}^{q-1}(k)$ within line 24; (h) $\hat{X}_{0}^{q-1}(k)$ in line 8 (subsequent iteration). The rectangular window is used to represent all spectra, so the effects of long-range spectral leakage are noticeable. Similarly, a 35\% harmonic distortion is selected for the same reason.}
\label{fig_TD_Spectrum}
\end{figure*}

The TD-IpDFT algorithm incorporating the OOBI compensation routine is summarized using the diagram in Fig. \ref{fig_FlowChart_Algo} and the pseudocode in Algorithm \ref{alg_TDIpDFT}, where the functions $\mathtt{TD\text{-}QSG}$, $\mathtt{TD\text{-}SR}$, $\mathtt{TD\text{-}APc}$ and $\mathtt{wf}$ refer, respectively, to Algorithm \ref{alg_TDQSG}, Algorithm \ref{alg_TDSR}, Algorithm \ref{alg_TDAPc} and (\ref{eq_rect_wind})-(\ref{eq_wf_def}), whereas Fig. \ref{fig_TD_Spectrum} shows the spectra at different steps of the process.
First, the delayed in-quadrature complex signal ($\bar{x}_{f}$) is generated using Algorithm \ref{alg_TDQSG} (line 1), followed by the calculation of its DFT spectrum and the application of the Hanning window in the frequency domain (lines 2-3). The IpDFT is then applied to obtain a first estimate of the signal parameters (line 4). Subsequently, an interference compensation loop is initiated (line 6) after the initialization of some auxiliary variables (line 5). These include the initial estimates of the positive ($\hat{X}_{i\text{+}}^{0}$) and negative ($\hat{X}_{i\text{--}}^{0}$) spectrums of a potential interference tone, the previous energy of the residual spectrum ($R_{e}^{0}$), along with the residual energy exit ($\tau_{R_{e}}$) and interference ($\tau_{i}$) trigger flags.

Within the loop, provided that the residual energy exit flag has not been triggered in the previous iteration, the contribution of the fundamental tone is first estimated ($\hat{X}_{0}^{q-1}$) by means of Algorithm \ref{alg_TDSR} (line 8). Algorithm \ref{alg_TDSR}, named Time-Delay Spectral Reconstruction (TD-SR), calculates the spectrum of any tone '$\alpha$' in a delayed in-quadrature complex signal ($\bar{x}_{f}$) given estimates of its frequency ($\hat{f}_{\alpha}$) and those of the uncorrected amplitude ($\hat{A}_{\alpha\text{+}}$) and phase ($\hat{\varphi}_{\alpha\text{+}}$) of its positive image. This is done by calculating the positive ($\sigma_{\text{+}_{\alpha}}$) and negative ($\sigma_{\text{--}_{\alpha}}$) complex delay gains (Algorithm \ref{alg_TDSR}: lines 1-2), which can then be used to obtain the spectral contributions of both positive and negative images (Algorithm \ref{alg_TDSR}: lines 3-6). Together with the previously estimated negative spectrum of the interference tone ($\hat{X}_{i\text{--}}^{q-1}$),  $\hat{X}_{0}^{q-1}$ is removed from the original spectrum (line 9) and used to determine whether such narrowband components may be present.
The analysis of such bins, which should correspond to a positive image of a spurious tone, is performed during the first iteration. Among them, only those where an interference tone is expected are considered, i.e., for a three-cycle window $k\in[0-2]\cup[4-7]$. Once the highest magnitude bin $k_{c}$ has been determined (\ref{eq_kc_range_def}), its total energy and that of its two closest neighbors are aggregated into $E_{c}$ (\ref{eq_Ec_def}) (line 11). The calculated $E_{c}$ is then compared with three heuristically defined threshold levels ($\lambda_{o}^{l}$, $\lambda_{o}^{u}$, $\lambda_{i}$), which allow for the identification of a potential interference (see Section \ref{doc_sec_4_sub2}). These thresholds measure the relation of $E_{c}$ to the total original spectral energy $E_{o}$ (\ref{eq_Eo_def}) ($\lambda_{o}^{l}$, $\lambda_{o}^{u}$) and to the entire set of interference bins $E_{i}$ (\ref{eq_Ei_def}) ($\lambda_{i}$), that is, $\forall k \in [0, K-1]$.
\begin{equation}
\label{eq_kc_range_def}
k_{c} =  \arg \max_{k}{|\hat{X}_{i\text{+}}(k)|}; \quad k \in [0, 2] \cup [4, 7]
\end{equation}
\begin{equation}
\label{eq_Ec_def}
E_{c} =  \sum_{k}|\hat{X}_{i\text{+}}(k)|^{2}; \;
\begin{cases}
k \in [0, 2], & \text{if } k_{c} = 0 \\
k \in [5, 7], & \text{if } k_{c} = 7 \\
k \in [k_{c}\pm 1], & \text{otherwise} \end{cases}
\end{equation}
For an efficient implementation, a routine stop criterion is introduced inspired by the one proposed in \cite{Frigo2019ReducedLekageSynPhEstimation}, where the incremental ratio between consecutive values of $E_{i}/E_{o}$ was evaluated.  Instead, the evolution of the total residual spectral energy $R_{e}$ is used. This is defined as the total remaining energy after subtracting both the estimated fundamental and the interfering tones inferred by the previous iteration, from the original spectrum (\ref{eq_Re_def}). The adopted metric is more computationally efficient than the one used in \cite{Frigo2019ReducedLekageSynPhEstimation} since it substitutes divisions with subtractions. If the variation falls below a predefined threshold level ($\lambda_{R_{e}}$), $\tau_{R_{e}}$ is set to zero to stop the iterative process (see Section \ref{doc_sec_4_sub2}) (lines 21-25).
\begin{subequations}\label{eq_Eo_Ei_def}
\begin{align}
E_{o} =  \sum_{k=0}^{K-1}|X_{f_{H}}(k)|^{2}\label{eq_Eo_def}\\
E_{i} =  \sum_{k=0}^{K-1}|\hat{X}_{i\text{+}}(k)|^{2}\label{eq_Ei_def}
\end{align}
\end{subequations}
\begin{equation}
\label{eq_Re_def}
R_{e}^{q} =  \sum_{k=0}^{K-1}|X_{f_{H}}(k) - \hat{X}_{0}^{q-1}(k) - \hat{X}_{i}^{q-1}(k)|^{2}
\end{equation}
The second IpDFT is then applied to estimate the parameters of the positive image of the interfering component (line 22), which are used to approximate the complete spectrum of the interfering tone $(\hat{X}_{i}^{q}(k))$(line 23). Finally, the last IpDFT can be applied to $X_{f_{H}}(k) - \hat{X}_{i}^{q}(k)$, obtaining an improved estimate of the fundamental {$\{\hat{f_{0}^{q}},\hat{A_{0\text{+}}^{q}},\hat{\varphi_{0\text{+}}^{q}}\}$} (line 24). The process is looped Q times or until $\tau_{R_{e}}$ is triggered. The final results, the initial estimate, if no interferences are found, or the latest, once $\tau_{R_{e}}$ is triggered or the maximum number of iterations is reached, are then corrected to account for the amplitude and phase alterations due to the use of the delayed in-quadrature complex signal ($\bar{x}_{f}$). If no interferences are found, the amplitude and phase corrections can be applied directly (line 26), since the fundmanetal positive complex delay gain ($\sigma_{\text{+}_{0}}^{q-1}$) is already obtained when Algorithm 3 is applied to estimate the fundamental spectrum (line 8). Otherwise, this is done by means of Algorithm \ref{alg_TDAPc} (lines 30 and 35), named Time-Delay Amplitude and Phase correction (TD-APc), which corrects the estimated amplitude and phase of the positive image of any tone '$\alpha$' in $\bar{x}_{f}$ given $\hat{f}_{\alpha}$, $\hat{A}_{\alpha\text{+}}$, and $\hat{\varphi}_{\alpha\text{+}}$. 
Finally, fundamental frequency estimations at two successive reporting times $(m\;\text{and}\;m-1)$ are used to calculate the Rate-Of-Change-Of-Frequency (ROCOF) at the reporting time $m$ with a first-order backward approximation of a first-order derivative.
\begin{equation}
\label{eq_rocof_def}
\hat{\dot{f}}_{0}(m) = \left( \hat{f}_{0}(m) - \hat{f}_{0}(m-1) \right) F_{r}
\end{equation}
$F_{r}$ denotes the reporting rate. 



\begin{algorithm}
\caption{TD-IpDFT Algorithm}
\label{alg_TDIpDFT}
\begin{algorithmic}[1]
\REQUIRE $[x(n)]$
\STATE $\{ \bar{x}_{f}(n)\} = \mathtt{TD\text{-}QSG}[x(n)]$
\STATE $X_{f}(k) = \mathtt{DFT}[\bar{x}_{f}(n)]$
\STATE $ X_{f_{H}}(k)  =  0.5X_{f}(k) -0.25(X_{f}(k-1) + X_{f}(k+1))$
\STATE $\{\hat{f}_{0}^{0},\hat{A}_{0\text{+}}^{0},\hat{\varphi}_{0\text{+}}^{0}\} = \mathtt{IpDFT}[X_{f_{H}}(k)]$
\STATE $\hat{X}_{i\text{+}}^{0} (k) = 0 ; \hat{X}_{i\text{--}}^{0}(k) = 0 ; R_{e}^{0} = 0; \tau_{R_{e}} = 1; \tau_{i} = 0$
\FOR{$q=1$ \TO $Q$}
    \IF{$\tau_{R_{e}} = 1$}
         \STATE $\{\hat{X}_{0}^{q-1}(k),\sigma_{\text{+}_{0}}^{q-1}\}=\mathtt{TD\text{-}SR}[\hat{f}_{0}^{q-1},\hat{A}_{0\text{+}}^{q-1},\hat{\varphi}_{0\text{+}}^{q-1}]$
         \STATE $\hat{X}_{i\text{+}}^{q}(k) = X_{f_{H}}(k)- \hat{X}_{0}^{q-1}(k) - \hat{X}_{i\text{--}}^{q-1}(k)$
        \IF{$\tau_{i} = 0$}
            \STATE Apply (\ref{eq_kc_range_def})-(\ref{eq_Ec_def})-(\ref{eq_Eo_Ei_def})
            \IF{$(\frac{E_{c}}{E_{o}}\!\in\![\lambda_{o}^{l}, \lambda_{o}^{u}] \, \AND \,\frac{E_{c}}{E_{i}}\geq\lambda_{i})  \,\OR\,  \frac{E_{c}}{E_{o}}\!>\!\lambda_{o}^{u}$}
                \STATE $\tau_{i} = 1$
            \ENDIF
        \ENDIF
        \IF{$\tau_{i} = 1$}
            \STATE $\hat{X}_{r}^{q}(k) = \hat{X}_{i\text{+}}^{q}(k) - \hat{X}_{i\text{+}}^{q-1}(k)$
            \STATE $R_{e}^{q} = \sum|\hat{X}_{r}^{q}(k)|^{2}$
            \IF{$|R_{e}^{q}-R_{e}^{q-1}|<\lambda_{R_{e}}$}
                \STATE $\tau_{R_{e}} = 0$
            \ENDIF
            \STATE $\{\hat{f}_{i}^{q},\hat{A}_{i\text{+}}^{q},\hat{\varphi}_{i\text{+}}^{q}\} = \mathtt{IpDFT}[\hat{X}_{i\text{+}}^{q}(k)]$
            \STATE $\{\hat{X}_{i}^{q}(k),\text{-},\hat{X}_{i\text{+}}^{q}(k),\hat{X}_{i\text{--}}^{q}(k)\} = \mathtt{TD\text{-}SR}[\hat{f}_{i}^{q},\hat{A}_{i\text{+}}^{q},\hat{\varphi}_{i\text{+}}^{q}]$
            \STATE $\{\hat{f}_{0}^{q},\hat{A}_{0\text{+}}^{q},\hat{\varphi}_{0\text{+}}^{q}\} = \mathtt{IpDFT}[X_{f_{H}}(k)-\hat{X}_{i}^{q}(k)]$
            \ELSE
            \STATE $\hat{\varphi}_{0}^{q} = \hat{\varphi}_{0\text{+}}^{q} - \angle\sigma_{\text{+}_{0}}^{q} ;\;\hat{A}_{0}^{q} = \frac{\hat{A}_{0\text{+}}^{q}}{|\sigma_{\text{+}_{0}}^{q}|}$
        \STATE \textbf{break}
        \ENDIF
        \ELSE
        \STATE $\{\hat{A}_{0}^{q},\hat{\varphi}_{0}^{q}\} = \mathtt{TD\text{-}APc}[\hat{f}_{0}^{q},\hat{A}_{0\text{+}}^{q},\hat{\varphi}_{0\text{+}}^{q}]$
        \STATE \textbf{break}
     \ENDIF
\ENDFOR
\IF{$q=Q$}
    \STATE $\{\hat{A}_{0}^{Q},\hat{\varphi}_{0}^{Q}\} = \mathtt{TD\text{-}APc}[\hat{f}_{0}^{Q},\hat{A}_{0\text{+}}^{Q},\hat{\varphi}_{0\text{+}}^{Q}]$
\ENDIF
\ENSURE $\{\hat{f}_{0},\hat{A}_{0},\hat{\varphi}_{0}\}$
\end{algorithmic}
\end{algorithm}



\begin{algorithm}
\caption{TD-SR Algorithm}
\label{alg_TDSR}
\begin{algorithmic}[1]
\REQUIRE $[\hat{f}_{\alpha},\hat{A}_{\alpha\text{+}},\hat{\varphi}_{\alpha\text{+}}]$
\STATE $\varphi_{d_{\alpha}}=2\pi\hat{f}_{\alpha} d_{f} T_{s}$
\STATE $\sigma_{\text{+}_{\alpha}}= 1+e^{j(\frac{\pi}{2} - \varphi_{d_{\alpha}})}; \sigma_{\text{--}_{\alpha}}= 1+e^{j(\frac{\pi}{2} + \varphi_{d_{\alpha}})}$
\STATE $\hat{\varphi}_{\alpha\text{--}} = -(\hat{\varphi}_{\alpha\text{+}} - \angle\sigma_{\text{+}_{\alpha}})+\angle\sigma_{\text{--}_{\alpha}}$
\STATE $\hat{A}_{\alpha\text{--}} = \hat{A}_{\alpha\text{+}} \frac{|\sigma_{\text{--}_{\alpha}}|}{|\sigma_{\text{+}_{\alpha}}|}$
\STATE $\hat{X}_{\alpha\text{+}}(k)= \mathtt{wf}[\hat{f}_{\alpha},\hat{A}_{\alpha\text{+}},\hat{\varphi}_{\alpha\text{+}}]; \hat{X}_{\alpha\text{--}}(k)= \mathtt{wf}[\text{--}\hat{f}_{\alpha},\hat{A}_{\alpha\text{--}},\hat{\varphi}_{\alpha\text{--}}]$
\STATE $\hat{X}_{\alpha}(k)= \hat{X}_{\alpha\text{+}}(k)+\hat{X}_{\alpha\text{--}}(k)$
\ENSURE $\{\hat{X}_{\alpha}(k), \sigma_{\text{+}_{\alpha}},\hat{X}_{\alpha\text{+}}(k),\hat{X}_{\alpha\text{--}}(k)\}$
\end{algorithmic}
\end{algorithm}



\begin{algorithm}
\caption{TD-APc Algorithm}
\label{alg_TDAPc}
\begin{algorithmic}[1]
\REQUIRE $[\hat{f}_{\alpha},\hat{A}_{\alpha\text{+}},\hat{\varphi}_{\alpha\text{+}}]$
\STATE $\varphi_{d_{\alpha}}=2\pi\hat{f}_{\alpha} d_{f} T_{s}; \; \sigma_{\text{+}_{\alpha}}= 1+e^{j(\frac{\pi}{2} - \varphi_{d_{\alpha}})};$
\STATE $\hat{\varphi}_{\alpha} = \hat{\varphi}_{\alpha\text{+}} - \angle\sigma_{\text{+}_{\alpha}} ;\;\hat{A}_{\alpha} = \frac{\hat{A}_{\alpha\text{+}}}{|\sigma_{\text{+}_{\alpha}}|}$
\ENSURE $\{\hat{A}_{\alpha}, \hat{\varphi}_{\alpha}\}$
\end{algorithmic}
\end{algorithm}

\subsection{Novel Trigger Mechanism for OOBI Compenstation}\label{doc_sec_4_sub2}
A novel trigger mechanism for the OOBI that builds on and expands the normalized spectral energy method proposed in \cite{Derviskadic2018i-IpDFTforSynchrophasorEstimation} is adopted. In \cite{Derviskadic2018i-IpDFTforSynchrophasorEstimation} an energy threshold capable of identifying interfering components with distortion equal to or greater than 10\% was adjusted based on a characterization of the energy content shown by each test condition specified in \cite{PMU_Measurement_60255-118-1-2018}. A first refinement is proposed on the basis of an improved selection of the DFT bins to calculate the spectral energy. Although all bins in the interference spectrum were considered in \cite{Derviskadic2018i-IpDFTforSynchrophasorEstimation}, only $k_{c}$, and its two closest neighbors, are now selected. The ratio between $E_{c}$ and $E_{o}$ is called \textit{spectral energy ratio}. As shown in \cite{Derviskadic2018i-IpDFTforSynchrophasorEstimation}, the amplitude and phase steps are the main limiting signal dynamics when setting an adequate energy threshold. However, their energy is spread across the whole spectrum, whereas, in the case of an interfering tone, it will be concentrated around its frequency. This is leveraged by introducing an additional metric to enhance the distinction between interfering tones and other spurious contributions of different origin. The so-called \textit{spectral energy concentration ratio} allows to measure the spread of spectral energy around $k_{c}$ by taking the relation between $E_{c}$ and $E_{i}$.

\begin{figure}[!t]
\centering
\includegraphics[width=\linewidth]{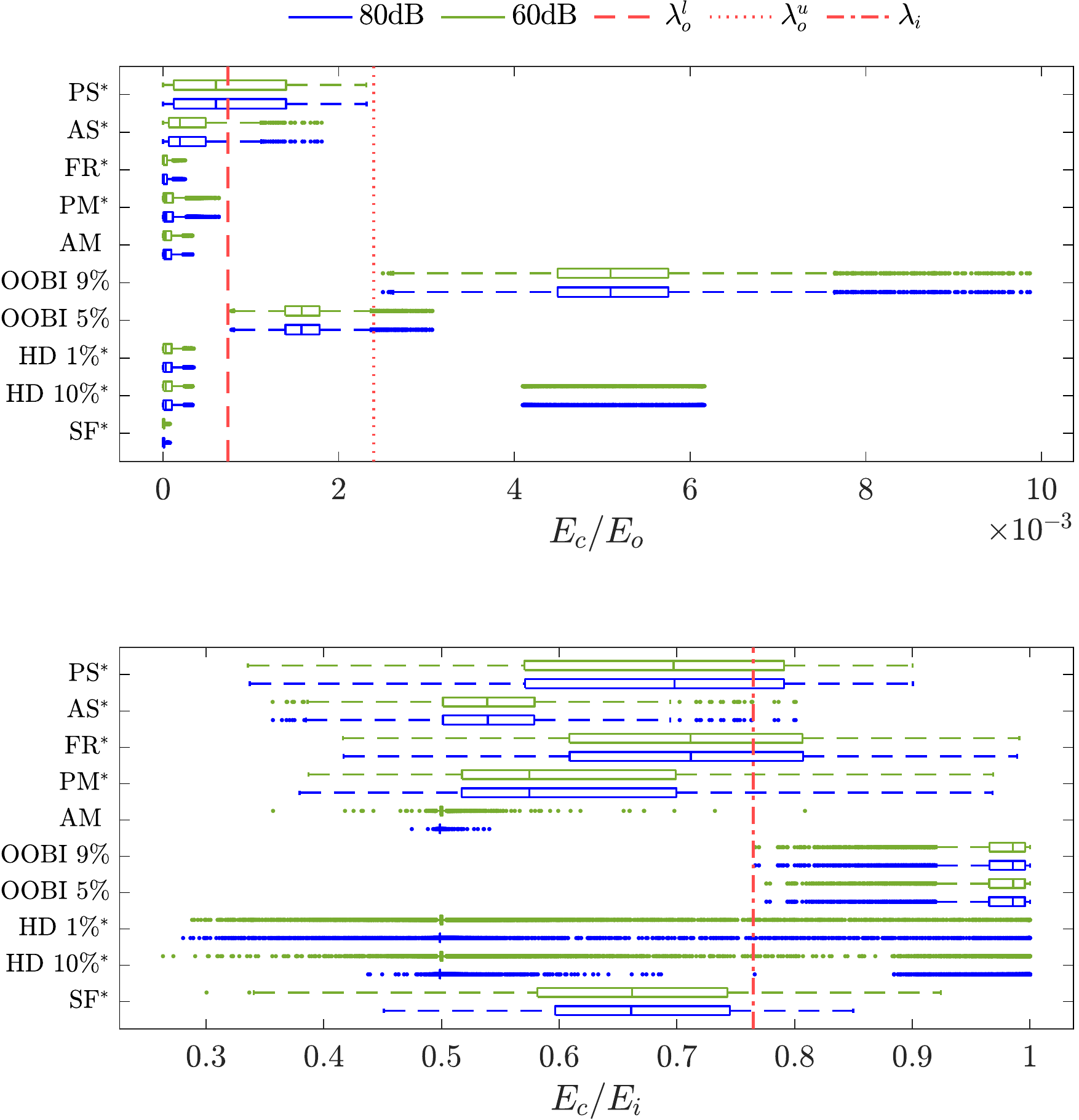}
\caption{Boxplot representation of $E_{c}/E_{o}$ (top) and $E_{c}/E_{i}$ (bottom). All operating conditions marked with $^{*}$ present a 10\% amplitude modulation with the highest modulating frequency within the range $[0.1-5]$ Hz compliant with \cite{PMU_Measurement_60255-118-1-2018}.}
\label{fig_OOBI_threshold}
\end{figure}

Setting appropriate values for $\lambda_{o}^{l}$, $\lambda_{o}^{u}$ and $\lambda_{i}$ allows for the identification and correction of OOBIs below the limit of 10\% set by \cite{PMU_Measurement_60255-118-1-2018}. To account for more realistic conditions and derive more robust thresholds, multi-dynamic signals, beyond the scope of \cite{PMU_Measurement_60255-118-1-2018}, have been simulated. Fig. \ref{fig_OOBI_threshold} shows the variability of $E_{c}/E_{o}$ and $E_{c}/E_{i}$ using a boxplot representation for each test condition in \cite{PMU_Measurement_60255-118-1-2018}. All regular tests have been combined with an amplitude modulation (except for the amplitude modulation itself and the OOBI) and evaluated for SNR equal to 60 and 80 dB. Amplitude modulations were selected because they present the second highest energy content, just after the steps. For each test, a 10\% amplitude modulation and the highest modulating frequency within the range $[0.1-5]$ Hz, compliant with \cite{PMU_Measurement_60255-118-1-2018}, has been used. Disregarding the amplitude modulated steps (PS* and AS*), Fig. \ref{fig_OOBI_threshold} (top) reveals how a minimum of 5\% OOBI can be distinguished from the other signal types by selecting a $\lambda_{o}^{l}= 7.4\cdot10^{-4}$. Lower levels of OOBI fall within the phase and amplitude modulated range (PM*). Additionally, the 10\% amplitude modulated harmonic distortion (HD 10\%*) also exhibits some outliers that exceed this limit. These correspond to the second harmonic which can also be corrected with the iterative process. Moreover, setting a $\lambda_{o}^{u}= 2.4\cdot10^{-3}$ allows distinguishing OOBIs with an amplitude greater than or equal to 9\% from amplitude modulated steps (PS* and AS*).

Regarding the value of $\lambda_{i}$, only an examination of the operating conditions that fall within the band defined by both $\lambda_{o}^{u}$ and $\lambda_{o}^{l}$ is necessary. Namely, the amplitude modulated steps. Therefore, a value of $0.765$ has been chosen, which allows to account for the \textit{spectral energy concentration ratio} of all OOBI distortion levels from 5\% to 9\%. Both OOBI cases are plotted (while the latter represents the most critical case) since the ratio decreases as the distortion level increases. A small overlap can still be seen in Fig. \ref{fig_OOBI_threshold} (bottom) regarding the PS* as well as some outliers for the AS*. However, the simulation results according to the standard \cite{PMU_Measurement_60255-118-1-2018} did not show any impact on their respective tests. All the cases (top and bottom) reveal that noise levels have no significant impact on the magnitude of neither $E_{c}/E_{o}$ nor $E_{c}/E_{i}$. Only for the latter in the signal frequency (SF*), amplitude modulation (AM*) and harmonic distortion (HD*) tests can the effects of noise be noticed. 

\begin{figure}[!t]
\centering
\includegraphics[width=\linewidth]{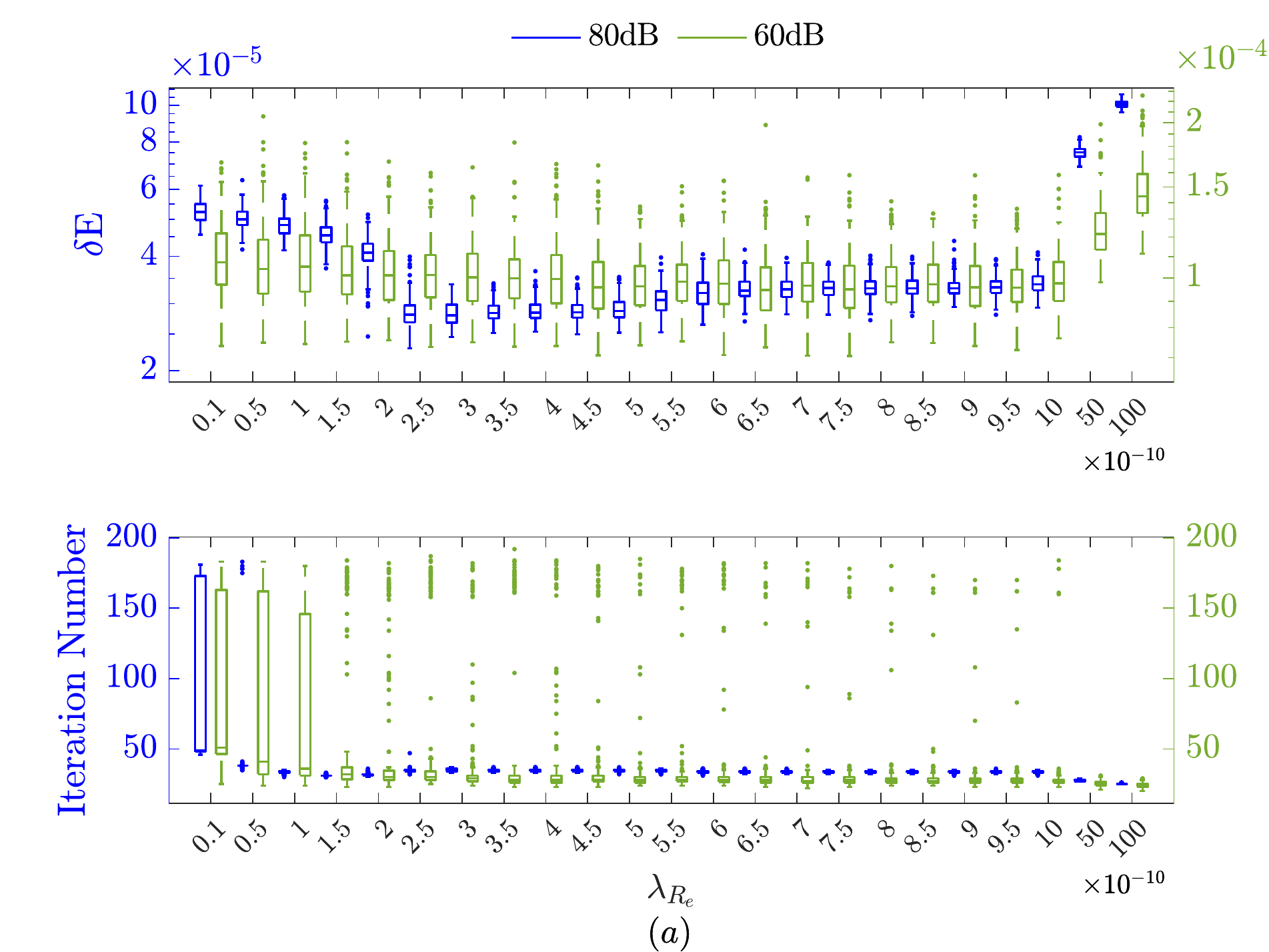}
\includegraphics[width=\linewidth]{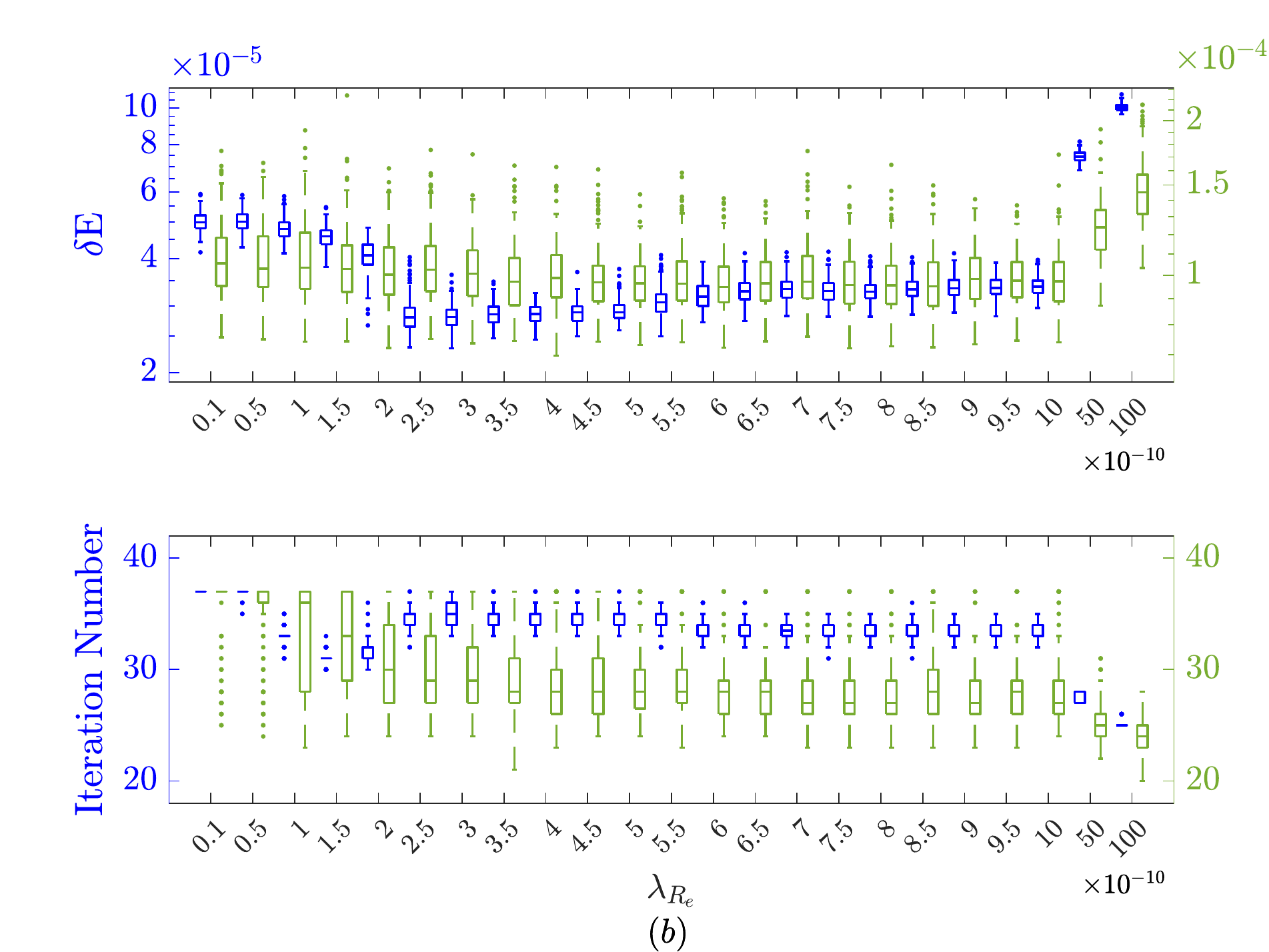}
\caption{Boxplot representation of the error in estimating the correction term $\delta$ (top) and the total number of iterations (bottom) for different values of $\lambda_{R_{e}}$. The maximum number of iterations Q is set to 200 (a) and 37 (b).}
\label{fig_OOBI_tuning}
\end{figure}

Finally, the adjustment of $\lambda_{R_{e}}$ described in Section \ref{doc_sec_4_sub1}, is analyzed. This is done by simulating 200 consecutive runs of a 10\% distortion OOBI test for 23 different potential threshold values. These have been selected within the $10^{-8}$ to $10^{-11}$ range based on a previous coarse estimation. Each run consists of 43 windows of analysis for each interfering tone. The analysis is done accounting for SNRs equal to 60 and 80 dB and allowing the iterative process to run, if unstopped, for a sizeable number of iterations.  Fig. \ref{fig_OOBI_tuning} shows the variability of the error in estimating the correction term $\delta$ and in the total number of iterations executed for a maximum Q equal to 200 (Fig. \ref{fig_OOBI_tuning}(a)) and 37 (Fig. \ref{fig_OOBI_tuning}(b)) . The error estimates of $\delta$ consider the global maximum error among all simulated interfering tones. Despite a noticeable performance trend in the error estimates of $\delta$ for the 80 dB case, a more uniform behavior is shown for a 60 dB noise. As shown in Fig. \ref{fig_OOBI_tuning}(a) for the 60dB case, although no single run reached the maximum iteration limit (Q) of 200, all considered threshold values $\lambda_{R_{e}}$, with the exception of $\lambda_{R_{e}} = 10^{-8}$ and $\lambda_{R_{e}} = 5\cdot10^{-9}$, presented runs where a large number of executed iterations were required. Although the adoption of $\lambda_{R_{e}} = 10^{-8}$ or $\lambda_{R_{e}} = 5\cdot10^{-9}$ would limit computational cost by reducing the maximum number of required iterations, it would also compromise accuracy, as both values result in notably higher error estimates of $\delta$ at a lower noise level. Fig. \ref{fig_OOBI_tuning}(b) examines the impact of limiting the maximum number of iterations (Q) to 37 on the error estimates of $\delta$\footnote{A value of $\text{Q} = 37$ has been selected as it corresponds to the maximum number of iterations required to attain the lowest maximum error in the estimates of $\delta$ within the 80dB case in Fig. \ref{fig_OOBI_tuning}(a).}. The results show that there are no significant changes in the error values obtained compared to Fig. \ref{fig_OOBI_tuning}(a). Thus, a $\lambda_{R_{e}} = 9.5\cdot10^{-10} $ is selected as a trade-off between accuracy and computational performance. This value ensures the lowest maximum error in the estimates of $\delta$ ($1.3683\cdot10^{-4}$ (60 dB)) when considering an absolute maximum of 37 iterations.

\subsection{Computational Complexity}\label{doc_sec_4_sub3}
To evaluate the feasibility of the implementation of the TD-IpDFT in an embedded device, and compare its performance with other state-of-the-art techniques, its computational complexity is here analyzed. Table \ref{tab_Complx} summarizes the total number of arithmetic operations required by the TD-IpDFT (both in case an interfering tone is detected or not) versus those used by its constituent functions, e.g., $\mathtt{IpDFT}$.
\begin{figure}[!t]
\centering
\includegraphics[width=\linewidth]{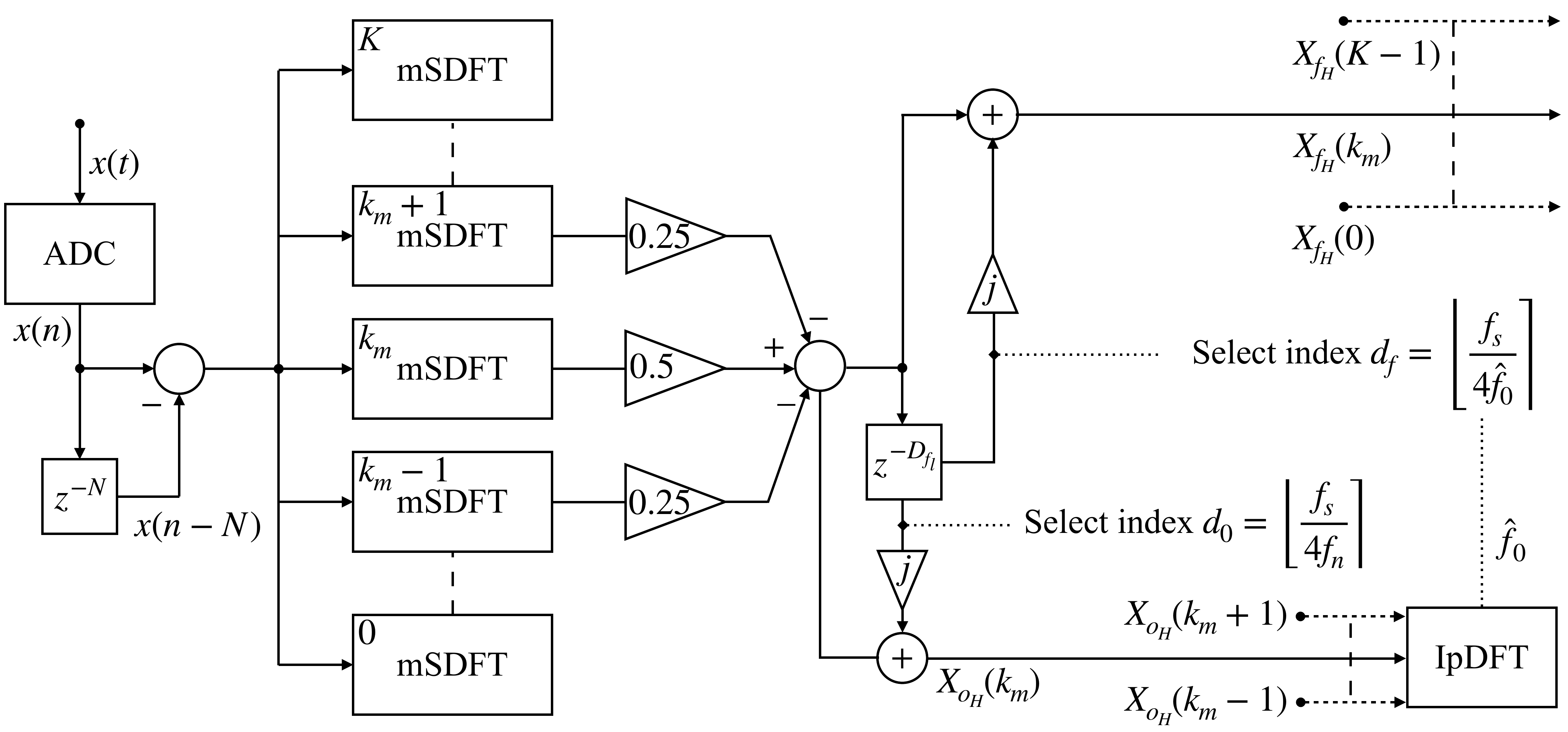}
\caption{Combined mSDFT and TD-QSG implementation.}
\label{fig_mSDFT_TDQSG}
\end{figure}
As in \cite{Derviskadic2018i-IpDFTforSynchrophasorEstimation} and \cite{Frigo2019ReducedLekageSynPhEstimation}, the difference between simple operations ($+|-|\times$), complex operations ($\div|\exp|\sin|\angle$), and function calls (such as calls to predefined subroutines or algorithms, e.g. $\mathtt{IpDFT}$) is drawn considering a field programmable gate array-based device (FPGA). Likewise, the total cost is expressed in terms of the total number of DFT bins calculated, $K$, and the maximum number of executions of the iterative process Q. The results show a total of 675 simple operations and 164 complex operations are required if no interference is detected and $174 + 1065\text{Q}$ simple and $34 + 287\text{Q}$ complex operations considering the OOBI iterative compensation for 8 DFT bins. For a maximum of 37 iterations the total number of operations results in 39579 simple and 10653 complex operations.
No explicit method has been specified for the calculation of the DFT bins, nor for generating the delayed in-quadrature complex signal ($\bar{x}_{f}(n)$). As indicated in \cite{Derviskadic2018i-IpDFTforSynchrophasorEstimation}, if a small number of DFT bins are needed, recursive computation methods are generally more efficient. Among them, the mSDFT technique \cite{Duda2010AccurateGuaranteedStableSlidingDFT} is guaranteed stable without sacrificing accuracy and has been used by previous implementations of IpDFT algorithms in embedded FPGA-based PMU devices \cite{Romano2014_mSDFT_eIpDFT}\cite{Derviskadic2020_iIpDFT_FPGA}. Consequently, Fig. \ref{fig_mSDFT_TDQSG} presents a preliminary implementation of the windowed delayed in-quadrature complex signal spectrum. For efficiency, as suggested in \cite{Duda2010AccurateGuaranteedStableSlidingDFT}, a common buffer is used for all mSDFT modules. Moreover, the delayed in-quadrature complex spectrum is obtained by buffering the windowed DFT bins of the real signal $X_{o_{H}}(k)$. As indicated in Algorithm \ref{alg_TDQSG} the method relies on an IpDFT to refine the final delay based on an initial nominal quadrature and thus select the appropriate past DFT bins from the buffer $D_{f_{l}}$. The size of the buffer $D_{f_{l}}$ depends on the lowest expected signal frequency, sampling rate, and the required number of DFT bins $K$. For a minimum frequency of 45 Hz, 50 kHz sampling rate, and 8 bins, its size will be roughly 1.5 times that of the one required by the mSDFT.


 

\begin{table}
\begin{center}
\begin{threeparttable}
\caption{TD-IpDFT Computational Complexity}
\label{tab_Complx}
\begin{tabular}{ l  c  c  c}
\toprule
& & Parameter & Value \\
\cmidrule{3-4}
& & q & $\leq$ Q\\
& & K & 8\\
\midrule
& $+|-|\times$ & $\div|\exp|\sin|\angle$ & $g(h)$ \\
\midrule
($h_{1}$) $\mathtt{IpDFT}$ (3p) & $15$ & $8$ & - \\
($h_{2}$) $\mathtt{wf}$ & $27\text{K}+6$ & $7\text{K}+6$ & - \\
($h_{3}$) $\mathtt{TD\text{-}QSG}$\tnote{1} & $10\text{K}+14$ & $5$ & - \\
($h_{4}$) $\mathtt{TD\text{-}SR}$ & $2\text{K}+17$ & $7$ & $2(h_{2})$ \\
($h_{5}$) $\mathtt{TD\text{-}APc}$ & $9$& $4$ & -\\
\midrule
Alg. II\tnote{1} & $+|-|\times$ & $\div|\exp|\sin|\angle$ & $g(h)$ \\
\midrule
$\mathtt{TD\text{-}IpDFT}$ (no int.) & $10\text{K}+9$ & $2\text{K}+4$ & $\sum_{i}^{1,3,4}(h_{i})$ \\
\\
& $8+\text{Q} +$ &$1+\text{Q} +$ & $(h_{3})+(h_{5})+$\\
$\mathtt{TD\text{-}IpDFT}$ (OOBI) & $(10\text{Q}+6)\text{K}$ & $(2\text{Q}+1)\text{K}$  & $2\text{Q}(h_{4}) + $\\
&  &  & $(2\text{Q}+1)(h_{1})$ \\
\bottomrule
\end{tabular}
\begin{tablenotes}
\item[1] For comparison with other existing methods the operations to obtain the delayed in-quadrature complex spectrum from the buffered windowed DFT bins of the
real signal, as shown in Fig. \ref{fig_mSDFT_TDQSG}, have been included.
\end{tablenotes}
\end{threeparttable}
\end{center}
\end{table}

\section{Performance Assessment}\label{doc_sec_5}
In this section, a complete evaluation of the TD-IpDFT algorithm is performed by comparing its performance with the static and dynamic accuracy limits indicated in \cite{PMU_Measurement_60255-118-1-2018} for the performance classes P and M. The assessment is carried out in a MATLAB simulation environment in terms of total vector error (TVE), frequency error (FE), ROCOF error (RFE) and response ($R_{t}$) and delay times ($D_{t}$) for the step tests.
As in \cite{Derviskadic2018i-IpDFTforSynchrophasorEstimation,Frigo2019ReducedLekageSynPhEstimation}, a nominal frequency of 50Hz and a reporting rate of 50 frames per second (fps) were selected to limit the number of tests. Similarly, the Hanning window function and a three-point interpolation scheme are adopted as they, respectively, represent a good trade-off between the width of the main lobe and the levels of the side lobes \cite{Grandke1983InterpolationAlgorithmsforDFTs}, and reduce the effect of long-range spectral leakage \cite{Agrez2002_WM_IpDFT}.
Each test is performed using a three nominal cycle window and considering two levels of additive white Gaussian noise. Noise levels with signal-to-noise ratios (SNR) equal to 60 and 80 dB have been selected, as previously used in \cite{Derviskadic2018i-IpDFTforSynchrophasorEstimation,Frigo2019ReducedLekageSynPhEstimation}, as they allow to consider the uncertainty of the measurement and simulate more realistic conditions. The results are presented by means of stacked graphs that summarize the resulting performance against the maximum permissible limits set by \cite{PMU_Measurement_60255-118-1-2018} (Figs. \ref{fig_SS_SF} - \ref{fig_DS_PS} ). Finally, the maximum values resulting from each test case are reported together again with the accuracy limits indicated in \cite{PMU_Measurement_60255-118-1-2018} in Tables \ref{tab_Static}, \ref{tab_Dynamic}, and \ref{tab_Steps}. All simulations are carried out according to the parameters given in Table \ref{tab_Param}.


 

\begin{table}
\begin{center}
\caption{TD-IpDFT Parameters}
\label{tab_Param}
\begin{tabular}{ c  c  c }
\toprule
Parameter & Variable & Value \\
\midrule
Nominal System Frequency & $f_{n}$ & 50 Hz\\
Window Type & -& Hann\\
Window Length & $T$ & 60 ms ($3/f_{n}$)\\
Sampling Rate & $F_{s}$ & 50 kHz \\
PMU Reporting Rate & $F_{r}$ & 50 fps\\
DFT bins & $K$ & 8\\
Max Number of Iterations & $Q$ & 37\\
IpDFT Interpolation Points & -& 3\\
Lower Spectral Energy Threshold & $\lambda_{o}^{l}$ & $7.4 \cdot 10^{-4}$\\
Upper Spectral Energy Threshold & $\lambda_{o}^{u}$ & $2.4 \cdot 10^{-3}$\\
Spectral Energy Concentration Threshold & $\lambda_{i}$ & $0.765$\\
Residual Energy Variation Threshold& $\lambda_{R_{e}}$ & $9.5 \cdot 10^{-10}$\\
\bottomrule
\end{tabular}
\end{center}
\end{table}


\subsection{Static Tests}
\begin{figure}[!t]
\centering
\includegraphics[width=\linewidth]{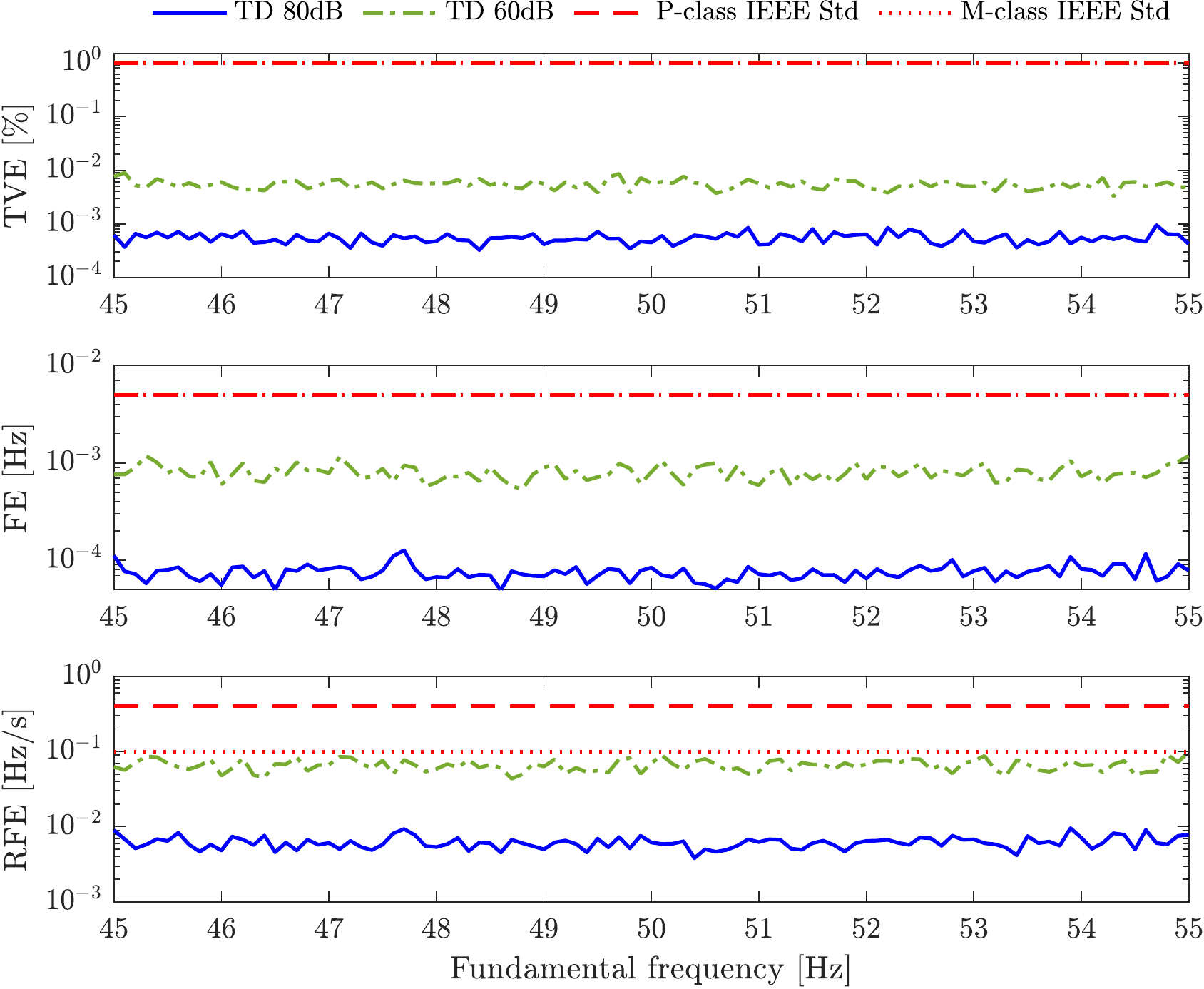}
\caption{Signal frequency range test \cite{PMU_Measurement_60255-118-1-2018}.}
\label{fig_SS_SF}
\end{figure}

\begin{figure}[!t]
\centering
\includegraphics[width=\linewidth]{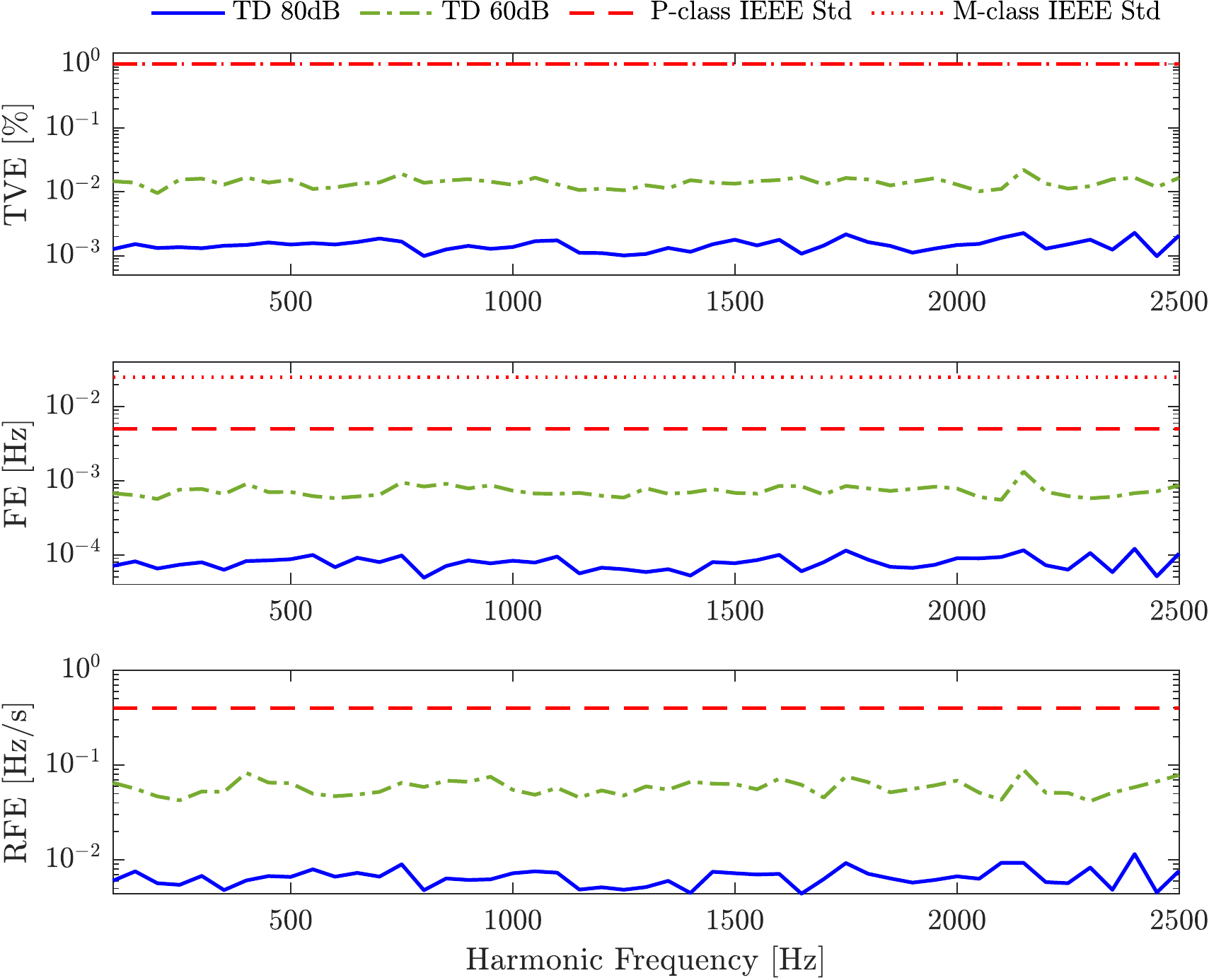}
\caption{Harmonic distortion test ($A_{h} = 10\% A_{0}$) \cite{PMU_Measurement_60255-118-1-2018}.}
\label{fig_SS_HD10}
\end{figure}

\begin{figure}[!t]
\centering
\includegraphics[width=\linewidth]{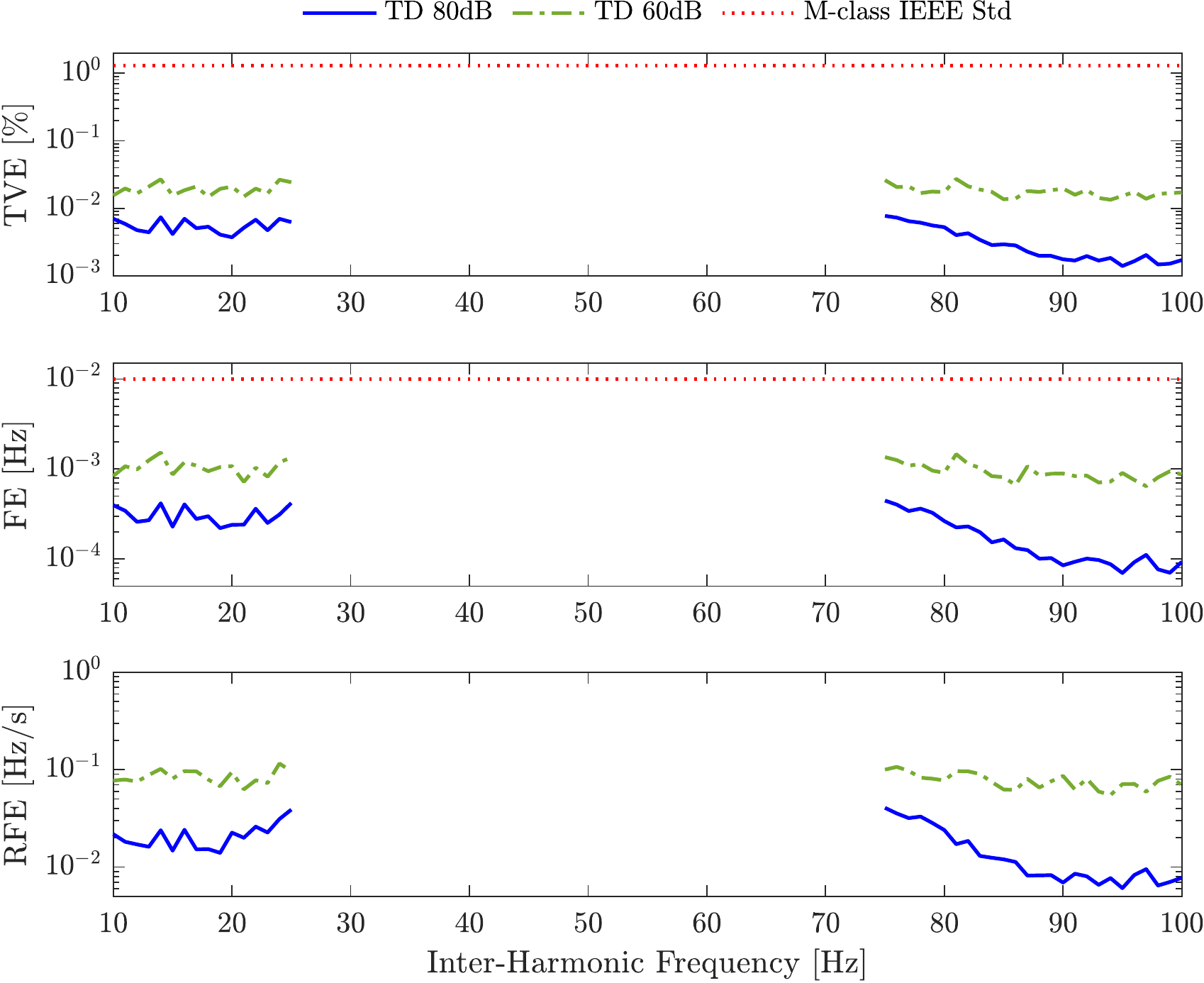}
\caption{OOBI test ($A_{ih} = 10\% A_{0}$).}
\label{fig_SS_OOBI}
\end{figure}

Three static tests are defined in \cite{PMU_Measurement_60255-118-1-2018} to evaluate the performance of the algorithm under steady-state conditions: the signal frequency range test, the harmonic distorsion test and the out-of-band interference (OOBI) test. The results of all static tests are presented in Figs. (\ref{fig_SS_SF} - \ref{fig_SS_OOBI}) and Table \ref{tab_Static}.

 

\begin{table*}
\begin{center}
\begin{threeparttable}
\caption{Maximum TVE, FE and RFE in Static Tests and Maximum Limit Allowed by \cite{PMU_Measurement_60255-118-1-2018}}
\label{tab_Static}
\begin{tabularx}{0.85\textwidth}{ >{\centering\arraybackslash}X  >{\centering\arraybackslash}X  c  c  c  c  c  c  c  c  c  c  c  c  c  c }
\toprule
& & \multicolumn{4}{c}{TVE[\%]} & & \multicolumn{4}{c}{FE[mHz]} & &\multicolumn{4}{c}{RFE[Hz/s]} \\
\cmidrule{3-6} \cmidrule{8-11} \cmidrule{13-16} 
& & \multicolumn{2}{c}{IEEE Std} & \multicolumn{2}{c}{TD-IpDFT}& &\multicolumn{2}{c}{IEEE Std} & \multicolumn{2}{c}{TD-IpDFT}& &\multicolumn{2}{c}{IEEE Std} & \multicolumn{2}{c}{TD-IpDFT}\\
& & P & M & \multicolumn{2}{c}{Hann $(3/f_{n})$} & & P & M & \multicolumn{2}{c}{Hann $(3/f_{n})$}& &P & M & \multicolumn{2}{c}{Hann $(3/f_{n})$}\\
\multicolumn{2}{c}{SNR [dB]} &  &  & 60 & 80 & &  & &60& 80& & & & 60& 80 \\
\midrule
\multicolumn{2}{c}{Sign Freq} & 1 & 1 & 0.009 & 0.001 & & 5 & 5 & 1.15 & 0.13 & & 0.4 & 0.1 & 0.099 & 0.010\\
\midrule
\multicolumn{2}{c}{Harm Dist 1\%} & 1 & 1 & 0.024 & 0.002 & & 5 & 25& 1.09 & 0.10 & & 0.4 & - &0.094 & 0.008\\
\multicolumn{2}{c}{Harm Dist 10\%} & 1 & 1 & 0.022 & 0.002 & & 5 & 25& 0.95 & 0.11 & & 0.4 & - & 0.089& 0.012\\
\midrule
 & 47.5Hz & - & 1.3 & 0.027 & 0.008 & & - & 10& 1.35 & 0.40& & - & - &0.102 & 0.041\\
 OOBI 10\%& 50Hz & - & 1.3 & 0.027 & 0.005 & & - & 10& 1.25& 0.28& & - & - & 0.116& 0.022\\
 & 52.5Hz & - & 1.3 & 0.027 & 0.007 & & - & 10& 1.36& 0.36& & - & - & 0.107& 0.036\\
 \midrule
 & 47.5Hz & - & 1.3 & 0.025& 0.009& & - & 10& 1.44& 0.39& & - & - & 0.110& 0.045\\
 OOBI 5\%\tnote{1} & 50Hz & - & 1.3 & 0.027& 0.005& & - & 10& 1.11& 0.26& & - & - & 0.103 & 0.023\\
 & 52.5Hz & - & 1.3 & 0.023 & 0.008& & - & 10& 1.18& 0.38& & - & -& 0.107& 0.036\\
 \bottomrule
\end{tabularx}
\begin{tablenotes}
\item[1] Maximum limit values taken from \cite{PMU_Measurement_60255-118-1-2018} for the 10\% case.
\end{tablenotes}
\end{threeparttable}
\end{center}
\end{table*}


The results of the signal frequency range test (Fig. \ref{fig_SS_SF}) show how the accuracy of the method does not depend on the fundamental tone frequency but rather on the total noise level, resulting in errors of one higher order of magnitude for the lower SNR. As reported in Table \ref{tab_Static}, maximum TVE values of 0.009\% (60 dB) and 0.001\% (80 dB) are obtained, well below the required 1\% limit. Likewise, similar trends also apply for the FE and RFE maximum errors, where respective values of 1.15 mHz (60 dB) and 0.13 mHz (80 dB) for the frequency and 0.099 Hz/s (60 dB) and 0.010 Hz/s (80 dB) for the ROCOF are obtained, in accordance with the limit requirements of 5 mHz and 0.1 Hz/s. As was the case in \cite{Derviskadic2018i-IpDFTforSynchrophasorEstimation,Frigo2019ReducedLekageSynPhEstimation}, it is worth noting that, under the higher noise conditions, spurious RFE values can marginally exceed the most stringent limit of class M.

The results of the harmonic distortion test are shown in Fig. \ref{fig_SS_HD10} for a THD of 10\% and in Table \ref{tab_Static} for both THDs of 1\% and 10\%. Again, no significant performance difference is detected on the basis of the order of the interference tone, but rather on the total noise level.  The 1\% performance limit is far from the maximum TVE values reported for the 60 dB case of 0.024\% (1\% THD) and 0.022\% (10\% THD) in Table \ref{tab_Static}. Specifically, the maximum values of FE and RFE were 1.09 mHz (1\% THD) and 0.95 mHz (10\% THD) for the frequency and 0.094 Hz/s (1\% THD) and 0.089 Hz/s (10\% THD) for the ROCOF. All are within the most demanding 5 mHz and 0.4 Hz/s limit requirements for devices of class P.

Finally, regarding the OOBI test, Fig. \ref{fig_SS_OOBI} shows the maximum value of TVE, FE and RFE obtained for each interference tone among the three simulated fundamental frequency values of 47.5, 50 and 52.5 Hz considering a total interharmonic distortion of 10\% as required by \cite{PMU_Measurement_60255-118-1-2018}. The closer the interference is to the fundamental tone, the more difficult it is to detect and remove it. This trend is evident in Fig. \ref{fig_SS_OOBI} for the lower noise level, as the maximum registered values increase as they approach the fundamental tone frequency. These maximum values, together with those for a 5\% distortion, disaggregated by each fundamental frequency, are presented in Table \ref{tab_Static}. All values are well within the performance requirements of class M. It is important to note that no performance requirement is provided in \cite{PMU_Measurement_60255-118-1-2018} for interfering signals with a magnitude other than 10\% of the fundamental. Thus, the same reference values have been considered for the 5\% case. As shown, the newly proposed trigger mechanism for OOBI is capable of correcting interferences with amplitudes equal to or greater than 5\%, showing a very similar performance to the 10\% case. This represents an improvement over the detection mechanism used in \cite{Derviskadic2018i-IpDFTforSynchrophasorEstimation,Frigo2019ReducedLekageSynPhEstimation}, which was only able to detect and correct interferences equal to or greater than 10\%.

\subsection{Dynamic Tests}

 

\begin{table*}[!t]
\begin{center}
\caption{Maximum TVE, FE and RFE in Dynamic Tests and Maximum Limit Allowed by \cite{PMU_Measurement_60255-118-1-2018}}
\label{tab_Dynamic}
\begin{tabular}{ c  c  c  c  c  c  c  c  c  c  c  c  c  c  c c }
\toprule
& & \multicolumn{4}{c}{TVE[\%]} & & \multicolumn{4}{c}{FE[mHz]} & &\multicolumn{4}{c}{RFE[Hz/s]} \\
\cmidrule{3-6} \cmidrule{8-11} \cmidrule{13-16} 
& & \multicolumn{2}{c}{IEEE Std} & \multicolumn{2}{c}{TD-IpDFT}& &\multicolumn{2}{c}{IEEE Std} & \multicolumn{2}{c}{TD-IpDFT}& &\multicolumn{2}{c}{IEEE Std} & \multicolumn{2}{c}{TD-IpDFT}\\
& & P & M & \multicolumn{2}{c}{Hann $(3/f_{n})$} & & P & M & \multicolumn{2}{c}{Hann $(3/f_{n})$}& &P & M & \multicolumn{2}{c}{Hann $(3/f_{n})$}\\
\multicolumn{2}{c}{SNR [dB]} &  &  & 60 & 80 & &  & &60& 80& & & & 60& 80 \\
\midrule
\multicolumn{2}{c}{Ampl Mod} & 3 & 3 &0.649 & 0.646& & 60 & 300 & 1.04& 0.11& & 2.3 & 14 & 0.087& 0.009\\
\multicolumn{2}{c}{Ph Mod} & 3 & 3 & 0.563& 0.558& & 60 & 300& 19.26& 18.94& & 2.3 & 14 & 0.634 & 0.599\\
\midrule
\multicolumn{2}{c}{Freq Ramp} & 1 & 1 & 0.044& 0.038& & 10 & 10& 1.26& 0.12& & 0.4 & 0.2 & 0.104 &0.011\\
\bottomrule
\end{tabular}
\end{center}
\end{table*}

\begin{figure}[!t]
\centering
\includegraphics[width=\linewidth]{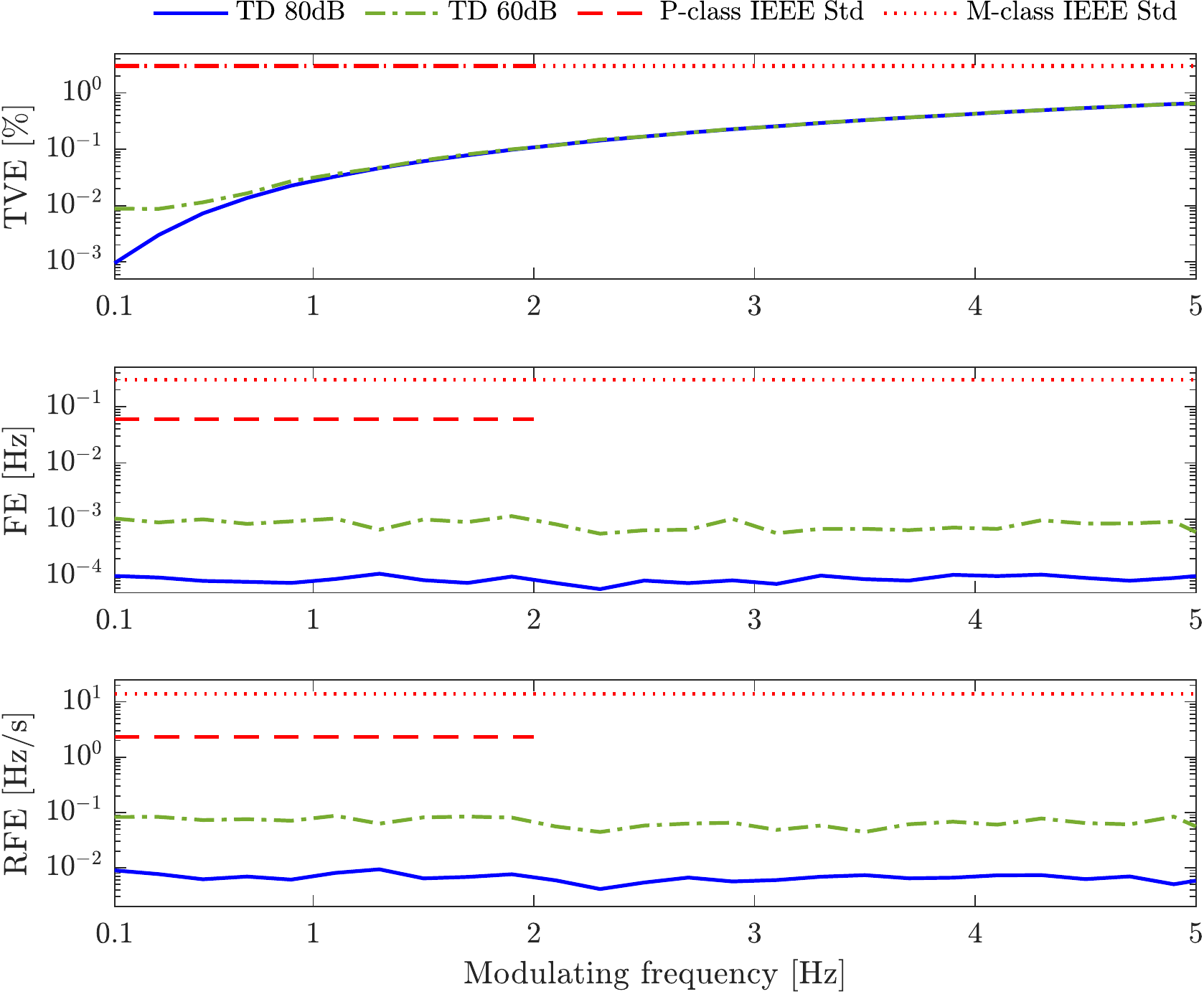}
\caption{Amplitude modulation test (depth 10\%).}
\label{fig_DS_AM}
\end{figure}

\begin{figure}[!t]
\centering
\includegraphics[width=\linewidth]{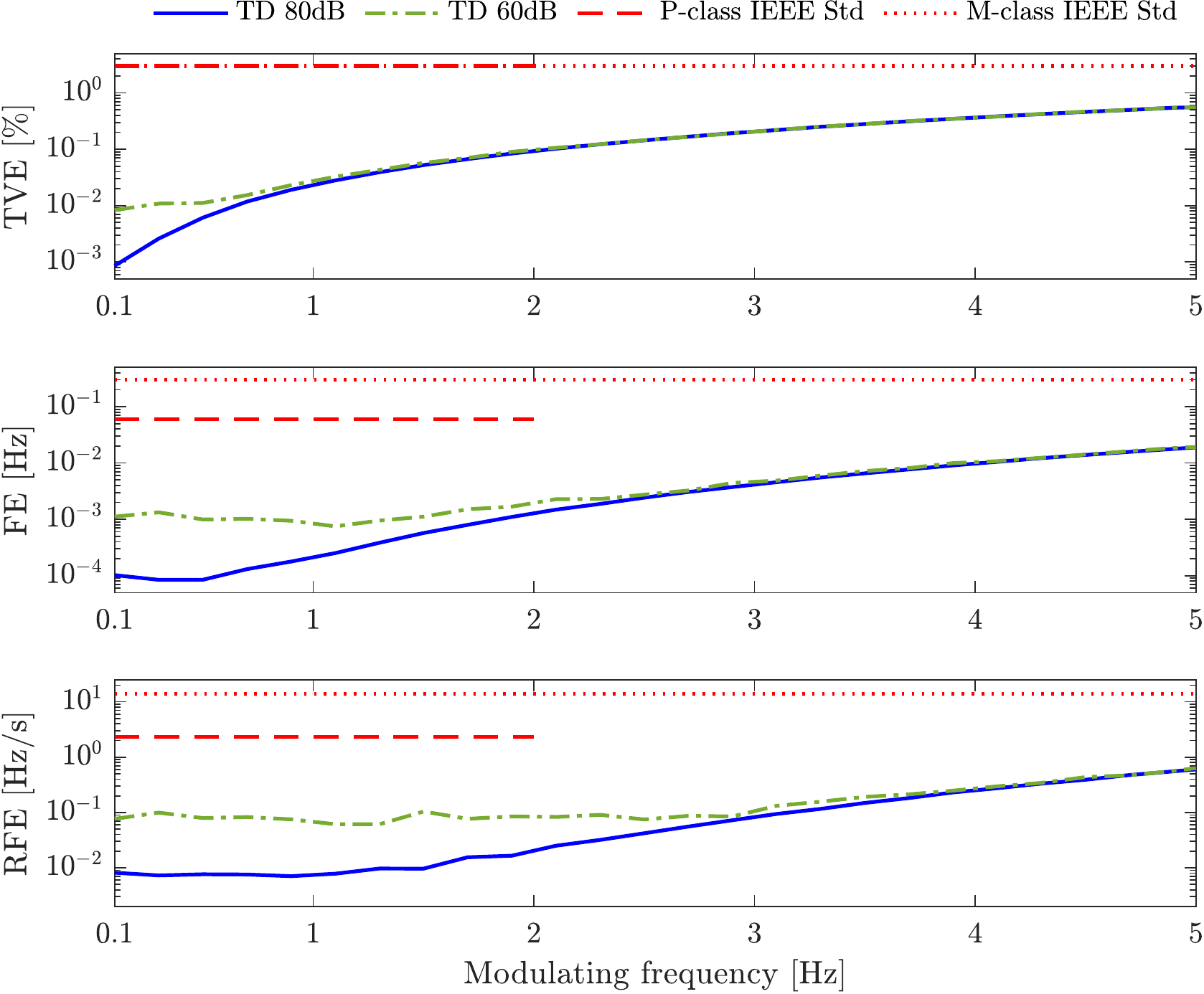}
\caption{Phase modulation test (depth $\pi/18$ rad).}
\label{fig_DS_PM}
\end{figure}

\begin{figure}[!t]
\centering
\includegraphics[width=\linewidth]{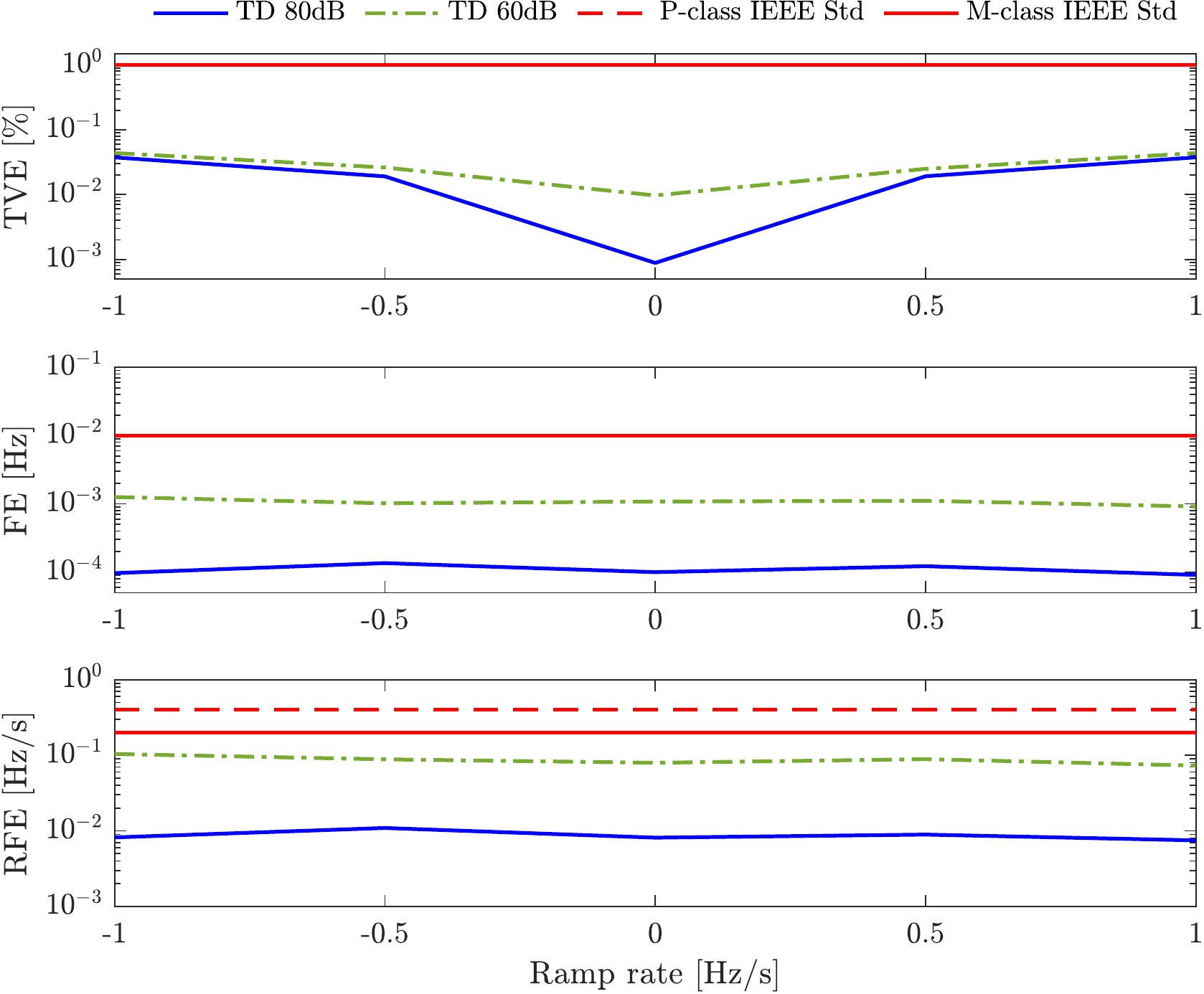}
\caption{Frequency ramp test. Worst-case TVE, FE, and RFE as functions of the ramp rate. \cite{PMU_Measurement_60255-118-1-2018}}
\label{fig_DS_FR}
\end{figure}

Two dynamic tests are defined in \cite{PMU_Measurement_60255-118-1-2018} to evaluate the performance of the algorithm under time-varying conditions: the measurement bandwidth test and the frequency ramp test. The results of all dynamic tests are presented in Figs. \ref{fig_DS_AM} - \ref{fig_DS_FR} and Table \ref{tab_Dynamic}.

The most demanding requirements set by \cite{PMU_Measurement_60255-118-1-2018} correspond to class P, with TVE, FE, and RFE values, respectively, of 3\%, 60 mHz, and 2.3 Hz/s for both amplitude and phase modulations. In the case of frequency ramps, class M sets the most stringent limits with TVE, FE, and RFE thresholds of 1\%, 10 mHz and 0.2 Hz/s, respectively.
As shown in Figs. \ref{fig_DS_AM} and \ref{fig_DS_PM}, both measurement bandwidth tests show how the algorithm exhibits similar TVEs. The total error increases with the modulation frequency and becomes the predominant source of error, masking the effect of noise for modulating frequencies above 1 Hz. However, the trends for the FE and RFE differ. On the one hand, in the phase modulation test (Fig. \ref{fig_DS_PM}), the TD-IpDFT method reveals a pattern like the one of the TVE where higher modulating frequencies result in higher errors, and noise as the main source of error is overridden. However, in the amplitude modulation test (Fig. \ref{fig_DS_AM}), the algorithm can provide accurate frequency and ROCOF estimates without being affected by the modulation frequency. All maximum errors fall well within the limits set in \cite{PMU_Measurement_60255-118-1-2018}.
The worst-case results of the frequency ramp test are shown in Fig. \ref{fig_DS_FR} for ramps with different positive and negative rates. The maximum values are presented in Table \ref{tab_Dynamic}. The TD-IpDFT is shown to provide accurate frequency and ROCOF estimates, in line with those of the signal frequency test and unaffected by the ramp rate, and maximum TVEs of 0.044\% (60 dB) and 0.038\% (80 dB) which depend on the magnitude of the ramp rate but not on its sign, fully meeting performance requirements.

\subsection{Step Tests}
\begin{figure}[!t]
\centering
\includegraphics[width=\linewidth]{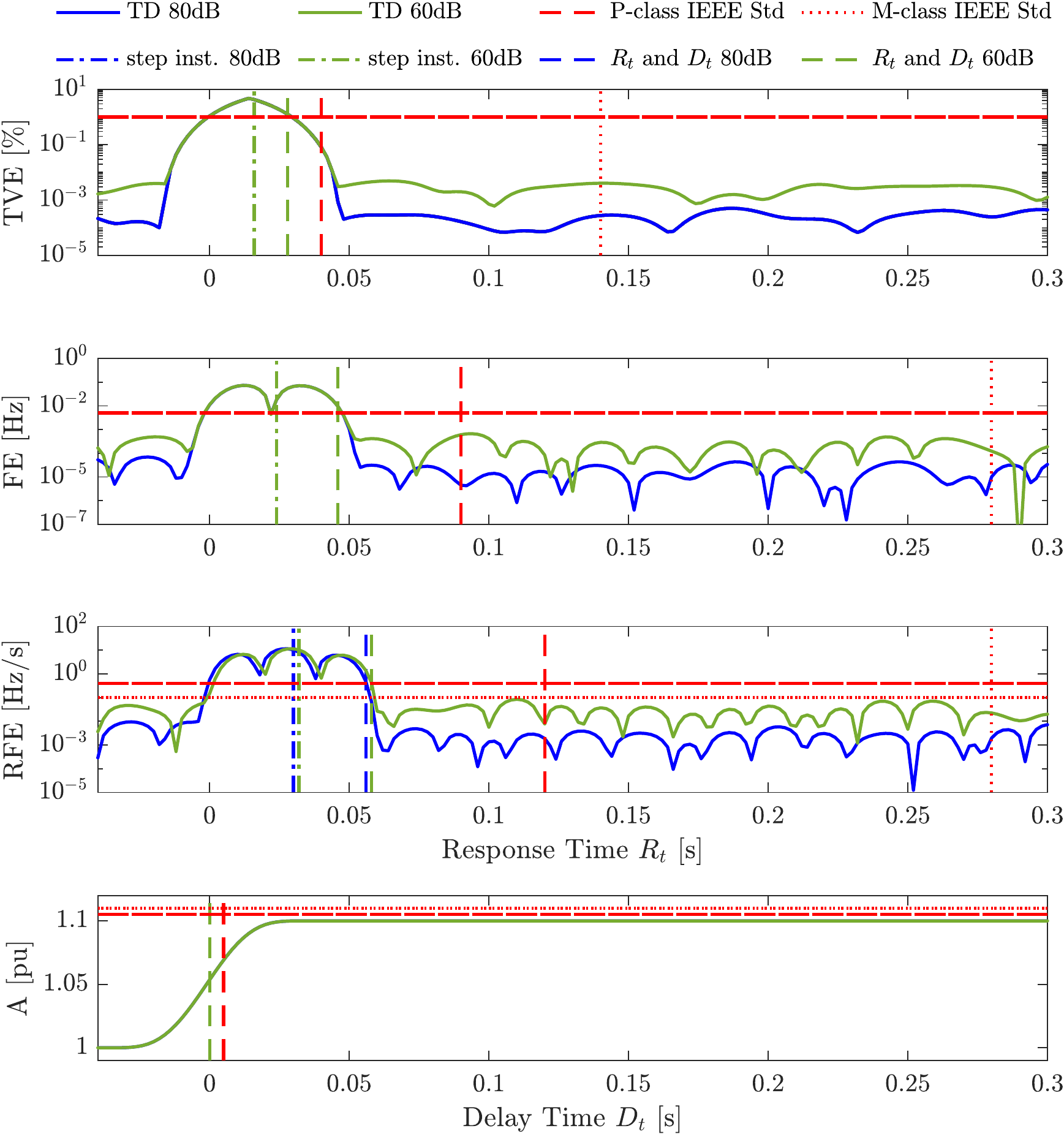}
\caption{Amplitude step test (+10\%) \cite{PMU_Measurement_60255-118-1-2018}.}
\label{fig_DS_AS}
\end{figure}

\begin{figure}[!t]
\centering
\includegraphics[width=\linewidth]{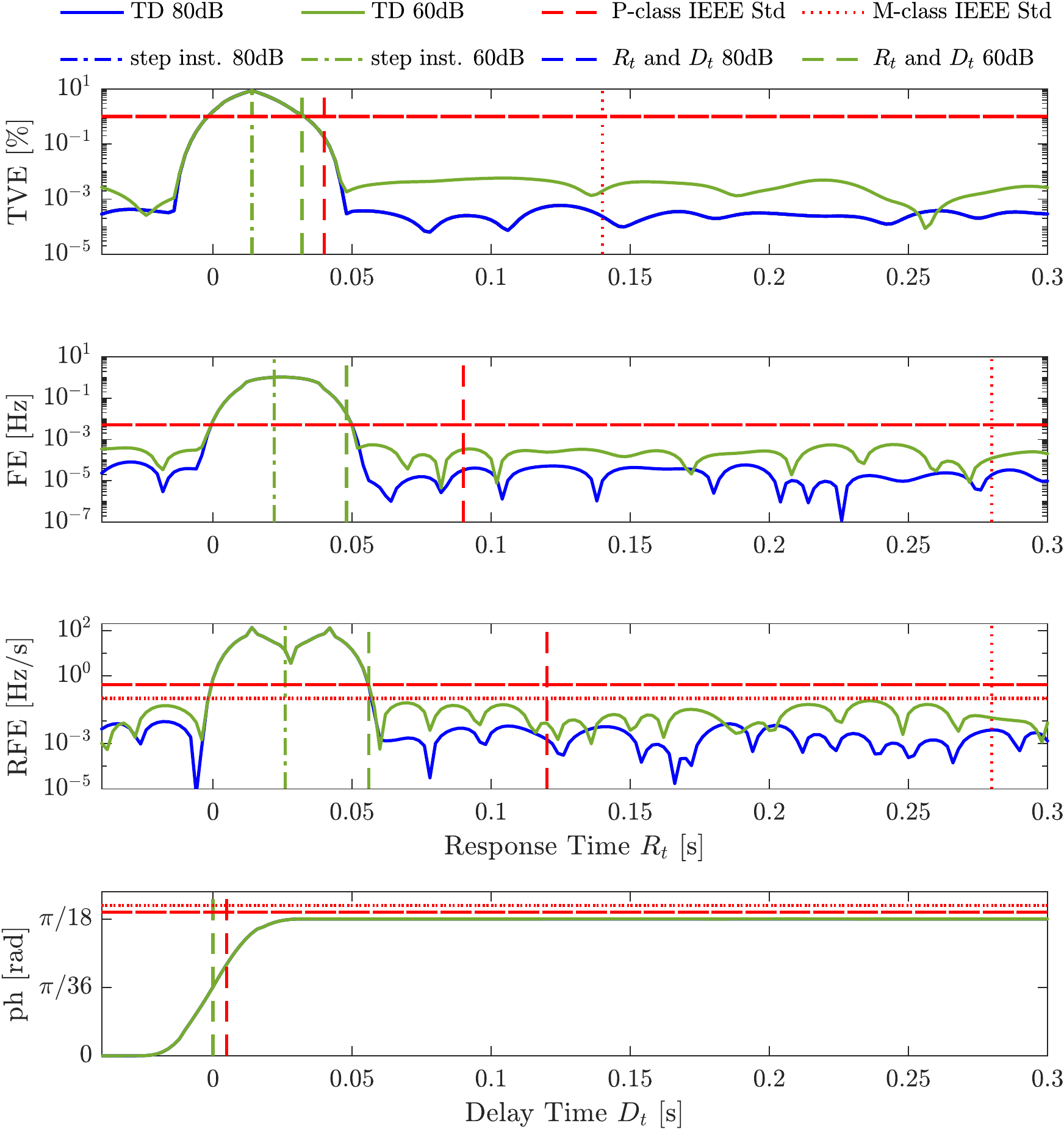}
\caption{Phase-step test (+$\pi$/18) \cite{PMU_Measurement_60255-118-1-2018}.}
\label{fig_DS_PS}
\end{figure}


 

\begin{table*}
\begin{center}
\caption{Maximum Response, Delay Times, and Overshoots in Step Tests and Limits Allowed by \cite{PMU_Measurement_60255-118-1-2018}}
\label{tab_Steps}
\begin{tabular}{ c  c  c  c  c  c  c  c  c  c  c  c  c  c  c c }
\toprule
& & \multicolumn{4}{c}{TVE Response Time [ms]} & & \multicolumn{4}{c}{FE Response Time [ms]} & &\multicolumn{4}{c}{RFE Response Time [ms]} \\
\cmidrule{3-6} \cmidrule{8-11} \cmidrule{13-16} 
& & \multicolumn{2}{c}{IEEE Std} & \multicolumn{2}{c}{TD-IpDFT}& &\multicolumn{2}{c}{IEEE Std} & \multicolumn{2}{c}{TD-IpDFT}& &\multicolumn{2}{c}{IEEE Std} & \multicolumn{2}{c}{TD-IpDFT}\\
& & P & M & \multicolumn{2}{c}{Hann $(3/f_{n})$} & & P & M & \multicolumn{2}{c}{Hann $(3/f_{n})$}& &P & M & \multicolumn{2}{c}{Hann $(3/f_{n})$}\\
\multicolumn{2}{c}{SNR [dB]} &  &  & 60 & 80 & &  & &60& 80& & & & 60& 80 \\
\midrule
\multicolumn{2}{c}{Ampl Step} & 40 & 140 & 28& 28& & 90 & 280 & 48& 48& & 120 & 280 & 60 & 60\\
\multicolumn{2}{c}{Ph Step} & 40 & 140 & 34& 34& & 90 & 280& 54& 54& & 120 & 280 & 60 & 60\\
\midrule
& & \multicolumn{4}{c}{Delay Time [ms]} & & \multicolumn{4}{c}{Max Overshoot [\%]}  \\
\cmidrule{3-6} \cmidrule{8-11} 
& & \multicolumn{2}{c}{IEEE Std} & \multicolumn{2}{c}{TD-IpDFT}& &\multicolumn{2}{c}{IEEE Std} & \multicolumn{2}{c}{TD-IpDFT}\\
& & P & M & \multicolumn{2}{c}{Hann $(3/f_{n})$} & & P & M & \multicolumn{2}{c}{Hann $(3/f_{n})$}\\
\multicolumn{2}{c}{SNR [dB]} &  &  & 60 & 80 & &  & &60& 80\\
\cmidrule{1-11}
\multicolumn{2}{c}{Ampl Step} & 5 & 5 & 4.5& 4.5& & 5 & 10 & 0& 0 \\
\multicolumn{2}{c}{Ph Step} & 5 & 5 & 4.5& 4.5& & 5 & 10& 0& 0 \\
\bottomrule
\end{tabular}
\end{center}
\end{table*}

Instantaneous changes in the amplitude ($\pm 10\%$) or phase ($\pm \pi/18$) of the signal are defined in \cite{PMU_Measurement_60255-118-1-2018} to evaluate the performance of the algorithm in a transient event.
The results of both steps are shown in Figs. \ref{fig_DS_AS} and \ref{fig_DS_PS}, and their response, delay times, and maximum overshoot values are presented in Table \ref{tab_Steps}. All results correspond to the positive step cases (similar results are obtained in the case of a negative step). Both Figs. \ref{fig_DS_AS} and \ref{fig_DS_PS} represent TVE, FE, and RFE as a function of their respective response times, which means that each time axis is centered at the moment when the accuracy limit is first exceeded. Likewise, the estimated phase and amplitude are shown as functions of their respective delay time, with their time axis centered at the instant each step occurs. The results show how the proposed algorithm meets all the requirements, with all estimates within the limits. It is important to note that no significant impact is found by the noise level during the transient, besides the pre- and post-steady-state accuracy already shown by the static tests.

\section{Comparison with SoTA IpDFT}\label{doc_sec_6}
This section is intended to discuss the results obtained with the TD-IpDFT compared to other IpDFT-based SE methods in the existing literature, namely the i-IpDFT \cite{Derviskadic2018i-IpDFTforSynchrophasorEstimation} and the HT-IpDFT \cite{Frigo2019ReducedLekageSynPhEstimation}, since all these SE methods have been designed to meet the accuracy requirements for the P and M performance classes.

\begin{figure}[!t]
\centering
\includegraphics[width=\linewidth]{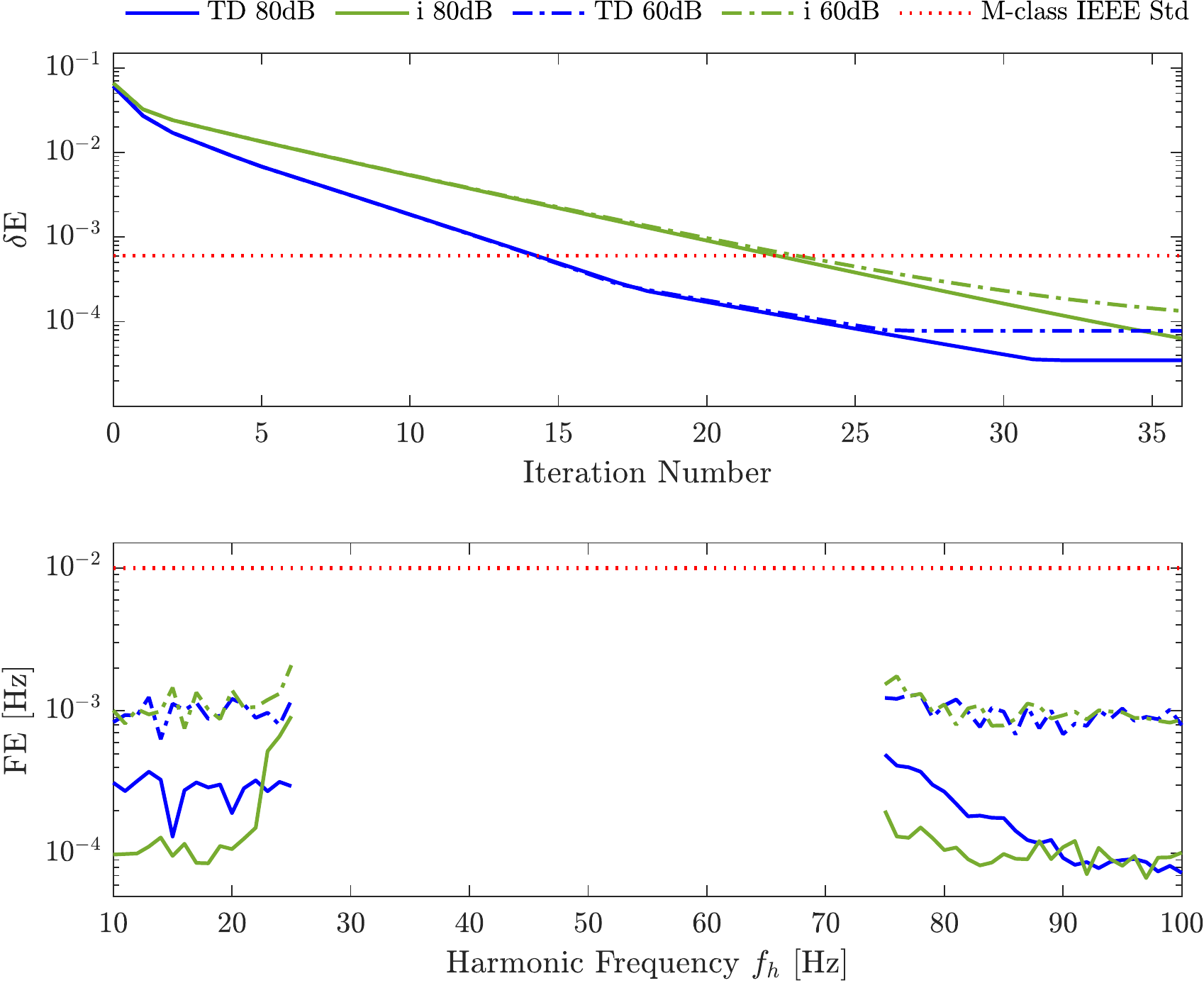}
\caption{Performance comparison between TD-IpDFT and i-IpDFT for noise levels with SNR equal to 60 and 80 dB. Error in estimating the correction term $\delta$ as a function of the iteration number (top) and FE as a function of the interference tone frequency for $Q = 37$ (bottom).}
\label{fig_Comp_OOBI}
\end{figure}

Focusing on the OOBI, which is the most computationally demanding test, the HT-IpDFT \cite{Frigo2019ReducedLekageSynPhEstimation} has been shown to offer a lower computational cost compared to the i-IpDFT \cite{Derviskadic2018i-IpDFTforSynchrophasorEstimation} due to the cancelation of negative frequency components offered by the adoption of the analytic signal. However, the HT-IpDFT does not meet the combined requirements of classes P and M for harmonic distortion (1\%) and phase step tests in \cite{PMU_Measurement_60255-118-1-2018}.
The TD-IpDFT satisfies all the accuracy requirements for the P and M classes, and also offers a reduction in the total computational cost compared to the i-IpDFT \cite{Derviskadic2018i-IpDFTforSynchrophasorEstimation,Derviskadic2020_iIpDFT_FPGA}. This is because it mitigates the effects of self-interference by using the in-quadrature complex signal for the estimate, while the i-IpDFT, which is based on the e-IpDFT \cite{Romano2014EnhancedInterpolatedDFT}, must iteratively compensate for them on each successive iteration\footnote{To draw a meaningful comparison a simplified version of Algorithm \ref{alg_TDIpDFT}, aimed at single-tone signals, can be defined consisting of lines 1-4 and line 30. This simplified TD-IpDFT can be shown to be more computationally efficient than the e-IpDFT. Taking into account the 3p variant of the e-IpDFT employed in the i-IpDFT and the same parameterization for both methods, the e-IpDFT requires 29K + 38 simple and 7K + 22 complex operations, while the simplified TD-IpDFT only requires 12K + 38 and 17 operations respectively.}. Taking into account the same parameterization for both methods, that is, the same type and length of the window and the same maximum number of iterations and DFT bins, the i-IpDFT requires a total of $844+1751\text{Q}$ simple operations and $218+474\text{Q}$ complex operations based on the updated tuning presented in \cite{Derviskadic2020_iIpDFT_FPGA}, whereas the TD-IpDFT can perform the estimation with just $174 + 1065\text{Q}$ and $34 + 287\text{Q}$ operations respectively. This represents approximately, for simple and complex operations, a total decrease of 40\% compared to the i-IpDFT. Moreover, considering the maximum overall error among the entire OOBI range, the TD-IpDFT also shows faster convergence, requiring 9 to 14 iterations less, for the same maximum error level compared to the i-IpDFT. This is shown in Fig. \ref{fig_Comp_OOBI} where a performance comparison between TD-IpDFT and i-IpDFT is presented for noise levels with SNR equal to 60 and 80 dB. Both the error in estimating the correction term $\delta$ as a function of the iteration number (top) and the FE obtained for the OOBI test considering $Q = 37$ (bottom) are shown. It can be seen that the TD-IpDFT requires fewer iterations to achieve the same maximum error level compared to the i-IpDFT indistinctly from the noise level. Furthermore, for $Q = 37$, the TD-IpDFT shows a modest performance improvement throughout the OOBI range for the high-noise case.  
Finally, the TD-IpDFT, as opposed to the i-IpDFT, also delivers a complex signal, which allows one to determine its envelope and angle. These could be used to implement identification and correction techniques for the amplitude and phase steps, as done in \cite{Karpilow2022_StepChangeDetec_for_RoCoF}.

\section{Conclusions}\label{doc_sec_7}
In this article, a SE technique based on the application of the IpDFT to a delayed in-quadrature complex signal has been presented. The use of the delayed in-quadrature complex signal has been found to be a simple and efficient way to mitigate the detrimental effects caused by the self-interference of the fundamental tone. This allows for a reduction in the total computational complexity of approximately 40\% compared to the i-IpDFT, which relies on an iterative correction based on the e-IpDFT. This is especially significant when the OOBI routine is triggered, since a faster convergence, requiring up to 14 fewer iterations for the same maximum error level, is obtained for the higher noise case.  However, an increase of around 150\% in memory requirements is necessary to generate the delayed signal.
The method represents an alternative to the i-IpDFT and the HT-IpDFT. Like the i-IpDFT, it satisfies all the accuracy requirements defined in the IEC/IEEE Std. 60255-118, while presenting a lower computational cost and it represents an easier and faster implementation than the cascaded Hilbert filter used by the HT-IpDFT. Furthermore, the novel OOBI trigger improves on the one used by the i-IpDFT and by the HT-IpDFT, which only allowed correction of distortions equal to or greater than 10\%. Future steps will include the implementation of the TD-IpDFT in an embedded device and its experimental validation.

\printbibliography 

\end{document}